\documentclass{tlp} 

\usepackage{latexsym} 
\usepackage{harvard} 

\newtheorem{examplerm}{Example}
\newenvironment{example}{\begin{examplerm}\rm}{\end{examplerm}}

\newtheorem{definitionrm}{Definition}
\newenvironment{definition}{\begin{definitionrm}\rm}{\end{definitionrm}}

\newtheorem{theorem}{Theorem}
\newtheorem{algorithmrm}{Algorithm}
\newtheorem{procedurerm}{Procedure}

\newcommand{\larr}{\leftarrow}
\newcommand{\NN}{\mbox{${I\!\!N}$}}

\def\mymath#1{\relax\ifmmode#1\else$#1$\fi}

\newenvironment{zitemize}
     {\begin{list}{--}{
     \setlength{\itemsep}{0 pt}
     \setlength{\parsep}{0 pt}
     \setlength{\topsep} {0 pt} }}
     {\end{list}}
\newenvironment{pitemize}
     {\begin{list}{}{
     \setlength{\itemsep}{0 pt}
     \setlength{\parsep}{0 pt}
     \setlength{\topsep} {2 pt} }}
     {\end{list}}
\newcommand{\homeo}{\unlhd}
\newcommand{\nothomeo}{\not\!\unlhd}
\newcommand{\homeostrict}{\lhd}

\newcommand{\ignore}[1]{} 
\newcommand{\forExperts}[1]{} 

\newcommand{\condtext}[2]{#1}

\newcommand{\dif}{\mathit{\backslash\!=}}

\newcommand{\pdset}[1]{{\cal #1}}

\newcommand{\ce}[2]{#1\circ #2}

\newcommand{\cpbeg}{\langle}
\newcommand{\cpdel}{,}
\newcommand{\cpend}{\rangle}

\newcommand{\cbox}{~\Box\,}

\begin{document}

\bibliographystyle{dcu} 

\title{Logic program specialisation through partial deduction: Control
issues} 
\author[M. Leuschel and M. Bruynooghe] 
{MICHAEL LEUSCHEL\\ 
Department of Electronics \& Computer Science,  
University of Southampton\\ Highfield, SO17 1BJ, UK  
\and 
 MAURICE BRUYNOOGHE\\ 
Department of Computer Science, Katholieke Universiteit Leuven\\ 
Celestijnenlaan 200A, B-3001 Heverlee, Belgium} 
\pagerange{\pageref{firstpage}--\pageref{lastpage}} 
\volume{\textbf{10} (3):}  
\jdate{September 2000} 
\setcounter{page}{1} 
\pubyear{2000} 

\maketitle 
\label{firstpage} 

\begin{abstract} 
  Program specialisation aims at improving the overall performance of
  programs by performing source to source transformations. A common
  approach within functional and logic programming, known respectively
  as partial evaluation and partial deduction, is to exploit partial
  knowledge about the input.  It is achieved through a well-automated
  application of parts of the Burstall-Darlington unfold/fold
  transformation framework. The main challenge in developing systems
  is to design automatic control that ensures correctness, efficiency,
  and termination.  This survey and tutorial presents the main
  developments in controlling partial deduction over the past 10 years
  and analyses their respective merits and shortcomings.  It ends with
  an assessment of current achievements and sketches some remaining
  research challenges.

\noindent 
{\bf Keywords:} program specialisation, logic programming, partial 
evaluation, partial deduction.
\end{abstract}

\section{Introduction} 
\label{section:basics:pe} 

Program specialisation aims at improving the overall performance of
programs by performing source to source transformations. A common
approach, known as {\em partial evaluation\/} is to guide the
transformation by partial knowledge about the input.  In contrast to
ordinary evaluation, partial evaluation is processing a given program
$P$ along with only {\em part\/} of its input, called the {\em static
  input}\index{static input}.  The remaining part of the input, called
the {\em dynamic input\/}\index{dynamic input}, will only be known at
some later point in time (which we call {\em runtime}).  Given the
static input $S$, the partial evaluator then produces a {\em
  specialised\/} version $P_S$ of $P$ which, when given the dynamic
input $D$, produces the same output as the original program $P$.  The
program $P_S$ is also called the {\em residual program}\index{residual
  program}.

The theoretical feasibility of this process, in the context of
recursive functions\index{recursive function}, has already been
established in \shortcite{Kleene:meta} and is known as Kleene's S-M-N
theorem.  However, while Kleene was concerned with theoretical issues
of computability and his construction often yields functions which are
more complex to evaluate than the original, the goal of partial
evaluation is to exploit the static input in order to derive more
efficient programs.

Consider, for example, the following program written in some informal
functional syntax, to compute the $n$-th power of a given value $x$,
where both $x,n\in\NN$.
\begin{example}\label{ex:power} 
\[\begin{array}{lll} 
\mathit{power}(x,n) & = & \mathit{if~} (n = 1) ~ \mathit{then~} x\\ 
& & \mathit{else~} (x * \mathit{power}(x,n-1))\\ 
\end{array}\] 
\end{example} 
Now, suppose we specialise the above program for the situation where
we want to compute the fifth power, that is $n=5$.  Looking at the
definition of the {\em power} function, we notice that the following
statements depend only on the value of $n$:
\begin{itemize} 
\item the test of conditional statement, 
\item the expression $n-1$ in the recursive call 
\item the recursive call, since the recursion is completely determined
  by the value of $n$.
\end{itemize}  
Performing these statements, and residualising the others, the result
of specialising the call {\em power(x,5)} is the residual program:
\[\begin{array}{lll} 
\mathit{power}(x,5) & = & x * x * x * x * x 
\end{array}\] 
If the specialiser is correct, the residual program computes the same
function as the original program, but naturally only for inputs of
which the static part equals the values with respect to which the
program was specialised. In the above example, the residual program
$\mathit{power}(x,5)$ still implements the {\em power} function, but
only the {\em fifth}-power function. It can be used to compute the
fifth power of any value, but can no longer compute the $n$-th power
of a value.

As the example illustrates, a partial evaluator evaluates those parts
of $P$ which only depend on the static input $S$ and generates code
for those parts of $P$ which require the dynamic input $D$. This
process has therefore also been called {\em mixed
  computation\/}\index{mixed computation} \cite{Ershov:TCS82}.  What
distinguishes partial evaluation from other program specialisation
approaches is that the transformation process is guided by the
available input.  Because part of the computation has already been
performed beforehand by the partial evaluator, the hope that we obtain
a more efficient program $P_S$ seems justified.
  
Partial evaluation
 \cite{ConselDanvy:POPL93,Jones:peval,Jones:ACMCompSurv96,%
MogensenSestoft:ECST97} has been applied to many programming 
 languages: functional programming languages (e.g., 
 \cite{Jones:peval}), logic programming languages (e.g., 
 \cite{Gallagher:PEPM93,komorowski:meta92,Pettorossi94:jlp}), 
\nocite{FujitaFurukawa88:ngc,FullerAbramsky:88,MogensenBondorf:LOPSTR92,%
Prestwich:LOPSTR92,BossiCoccoDulli:toplas90} functional logic
programming languages (e.g.,
\cite{AlpuenteFalaschiVidal:ESOP96,AlpuenteEtAl:TOPLAS98,%
AlbertEtAl:SAS98,LafaveGallagher:LOPSTR97}), term rewriting systems
(e.g., \cite{Bondorf:PEMC88,Bondorf:TAPSOFT89}), and imperative
programming languages (e.g., \cite{Andersen:CCC92,Andersen94:PhD}).
 
In the context of logic programming, full input to a program $P$
consists of a goal $G$ and evaluation corresponds to constructing a
complete SLDNF-tree for $P\cup\{G\}$.  For partial evaluation, the
static input takes the form of a goal $G'$ which is more general (less
instantiated) than a typical goal $G$ at runtime.  In contrast to
other programming languages, one can still execute $P$ for $G'$ and
(try to) construct an SLDNF-tree for $P\cup\{G'\}$.  So, at first
sight, it seems that partial evaluation for logic programs is almost
trivial and just corresponds to ordinary evaluation.
 
However, since $G'$ is not yet fully instantiated, the SLDNF-tree for
$P\cup\{G'\}$ is usually infinite and ordinary evaluation will not
terminate.  A technique which solves this problem is known under the
name of {\em partial deduction\/}\index{partial deduction}.  Its
general idea is to construct a finite number of finite trees and to
extract from these trees a new program that allows any instance of the
goal $G'$ to be executed.

\paragraph{Overview.} 
We will present the essentials of this technique in
Section~\ref{chapter:basics}.  Then, in
Section~\ref{section:basics:control} we identify the main issues in
controlling partial deduction, which we then address in much more
detail in Sections~\ref{section:basics:unfolding} and
\ref{section:global-control}.  In Section~\ref{section:conjPD} we then
discuss so-called {\em conjunctive partial deduction}, which extends
partial deduction in that it can specialise entire conjunctions
instead of just atoms.  Finally, in Section~\ref{section:discussion}
we discuss issues that arise for various extensions of logic
programming and conclude with a critical evaluation of the practical
applicability of existing partial deduction systems and techniques.

\paragraph{Terminology.} 
The term ``partial deduction'' has been introduced in
\cite{komorowski:meta92} to replace the term partial evaluation in the
context of pure logic programs (no side effects, no cuts).  Though in
Section~\ref{impure} we briefly touch upon the consequences of the
impure language constructs, we adhere to this terminology because the
word ``deduction'' places emphasis on the purely logical nature of the
source programs.  Also, while partial evaluation of functional and
imperative programs evaluates only those expressions which depend
exclusively on the static input, in logic programming one can, as we
have seen above, in principle also evaluate expressions which depend
on the unknown dynamic input.  This puts partial deduction closer to
techniques such as {\em unfold/fold\/}\index{unfold/fold} program
transformations \cite{BurstallDarlington:ACM77,Pettorossi94:jlp}, and
therefore using a different denomination seems justified.  Note that
partial evaluation and in particular partial deduction is not limited
to evaluation of expressions based on the static input. It can also
exploit data present in the source code of the program or gathered
though program analysis.  Finally, in the remainder of this article we
suppose familiarity with basic notions in logic programming
\cite{Apt90:htcs,Lloyd:flp}.


\section{Basics of Partial Deduction} 
\label{chapter:basics} 
 
In this section we present the technique of partial deduction, which
originates from \cite{Komorowski:POPL82}.  Other introductions to
partial deduction can be found in
\cite{komorowski:meta92,Gallagher:PEPM93,Leuschel:PE98a}.  Note that,
for clarity's sake, we deviate slightly from the original formulation
of \cite{Lloyd:jlp91}.

In order to avoid constructing infinite SLDNF-trees for partially
instantiated goals, the technique of {\em partial deduction\/} is
based on constructing finite, but possibly {\em incomplete\/}
SLDNF-trees.  The clauses of the specialised program are then
extracted from these trees by constructing one specialised clause per
branch.  A single resolution step with a specialised clause now
corresponds to performing {\em all\/} the resolutions steps (using
original program clauses) on the associated branch.
 
Before formalising the notion of partial deduction, we briefly recall
some basics of logic programming \cite{Apt90:htcs,Lloyd:flp}.
Syntactically, programs are built from an alphabet of variables (as
usual in logic programming, variable names start with a capital),
function symbols (including constants) and predicate symbols. Terms
are inductively defined over the variables and the function symbols.
Formulas of the form $p(t_1,\ldots,t_n)$ with $p/n$ a predicate symbol
of arity $n\geq 0$ and $t_1,\ldots, t_n$ terms are atoms. Literals
come in two kinds; positive literals are simply atoms; negative
literals are of the form $\mathit{not\ } A$ with $A$ an atom. A {\em
  definite clause\/} is of the form $ a \leftarrow B$ where the head
$a$ is an atom and the body $B$ is a conjunction of atoms. In {\em
  normal clauses}, the body $B$ is a conjunction of literals.  A
formula of the form $\leftarrow B$ with $B$ a conjunction of atoms is
a {\em definite goal\/}, with $B$ a conjunction of literals, it is a
{\em normal goal\/}. Definite, respectively normal {\em programs\/}
are sets composed of definite, respectively normal clauses.  In
analogy with terminology from other programming languages, a literal
in a clause body or in a goal is sometimes referred to as a {\em
  call}.

As detailed in \cite{Apt90:htcs,Lloyd:flp} a {\em derivation step\/}
selects an atom in a definite goal according to some {\em selection
  rule\/}.  Using a program clause, it first renames apart the program
clause to avoid variable clashes and then computes a most general
unifier ({\em mgu\/}) between the selected atom and the clause head
and, if an {\em mgu\/} exists, derives the {\em resolvent\/}, a new
definite goal.  (We also say that the selected atom is {\em
  resolved\/} with the program clause.)  Now, we are ready to
introduce our notion of SLD-derivation. As common in works on partial
deduction, it differs from the standard notion in logic programming
theory by allowing a derivation that ends in a nonempty goal where no
atom is selected.

\begin{definition}\label{def:incomplete-SLD-deriv} 
  Let $P$ be a definite program and $G$ a definite goal. An {\bf
    SLD-derivation} for $P \cup \{G\}$ consists of a possibly infinite
  sequence $G_0=G$, $G_1$, \ldots of goals, a sequence $C_1$, $C_2$,
  \ldots of properly renamed clauses of $P$ and a sequence $\theta_1$,
  $\theta_2$, \ldots of mgus such that each $G_{i+1}$ is derived from
  $G_i$ and $C_{i+1}$ using $\theta_{i+1}$.
\end{definition}

The initial goal of an SLD-derivation is also called the {\em query}.
An SLD-derivation is a successful derivation or refutation if it ends
in the empty clause, a failing derivation if it ends in a goal with a
selected atom that does not unify with any properly renamed clause
head, an incomplete derivation if it ends in a nonempty goal without
selected atom; if none of these, it is an infinite derivation. In
examples, to distinguish an incomplete derivation from a failing one,
we will extend the sequence of a failing derivation with the atom {\bf
  fail}.  The totality of SLD-derivations form a search space. One way
to organise this search space is to structure it in an SLD-tree. The
root is the initial goal; the children of a (non-failing) node are the
resolvents obtained by selecting an atom and performing all possible
derivation steps (a process that we call the {\em
  unfolding\/}\index{unfolding} of the selected atom). Each branch of
the tree represents an SLD-derivation. A {\em trivial\/} tree is a
tree consisting of a single node ---the root--- without selected atom.

SLDNF-derivations and SLDNF-trees originates from the extension
towards normal programs \cite{Apt94-1:jlp,Lloyd:flp}. As detailed in
\cite{Apt94-1:jlp}, a negative ground literal ${\mathit not}~A$ can be
selected in a goal, in which case a {\em subsidiary} SLDNF-tree is
built for the goal $\leftarrow A$. Eventually that tree contains a
refutation in which case the original goal fails, or fails finitely in
which case the original goal has a child ---the resolvent---
obtained by removing the negative literal (the mgu of this derivation
step is the empty substitution). Although it is possible that a
subsidiary tree never reaches the status where it contains a
refutation or fails finitely, we will ignore that possibility for the
time being, making the assumption that in such case the negative
literal is not selected and the subsidiary tree is not created (all
goals on branches extending the original goal will contain the
negative literal). This assumption, that we reconsider in Section
\ref{sec:subsidiary}, makes that the specialised program can be
extracted from the {\em main\/} tree, the tree that starts from the
initial goal. As a consequence, partial deduction for normal programs
is hardly different from partial deduction for definite programs.
Finally, we say that an SLDNF-tree (resp.\ SLDNF-derivation) is finite
iff the main tree (resp.\ derivation) is finite.  Observe that an
SLDNF-tree can be finite (and its construction can terminate) while
some of its subsidiary trees are infinite. Indeed, finding one
successful derivation in an infinite subsidiary tree is sufficient to
infer failure of the node containing the selected negative literal
referred to by the subsidiary tree.

Note that {\em floundering\/}, the situation where it is impossible to
select a literal in a goal because it consists solely of nonground
negative literals, is only a special case of an incomplete derivation.
In what follows, when we mention the branches of an SLDNF-tree, we
mean the branches of the main tree.

We now examine how specialised clauses can be extracted from
SLDNF-derivations and trees.

\begin{definition}\label{def:basics:resultant-derivation} 
  Let $P$ be a program, $G = \larr Q$ a goal, $D$ a finite
  SLDNF-derivation of $P\cup\{G\}$ ending in $\larr B$, and $\theta$
  the composition of the {\em mgu}s in the derivation steps.
  Then the formula $Q\theta \larr B$ is called the {\bf
    resultant\/}\index{resultant} of $D$.
\end{definition} 

Note that the formula is a clause when $Q$ is a single atom, as is the
case in standard partial deduction. {\em Conjunctive partial
  deduction\/} (Section \ref{section:conjPD}) also allows $Q$ to be a
conjunction of several atoms.  The relevant information to be
extracted from an SLDNF-tree is the set of resolvents and the set of
atoms occurring in the literals at the non-failing leaves.

\begin{definition} \label{def:resultants} 
  Let $P$ be a program, $G$ a goal, and $\tau$ a finite SLDNF-tree for
  $P\cup\{G\}$.
  Let $D_1,\ldots,D_n$ be the non-failing SLDNF-derivations associated
  with the branches of $\tau$.
  Then the {\bf set of resultants}, $resultants(\tau)$, is the set
  whose elements are the resultants of $D_{1},\ldots,D_{n}$ and the
  {\bf set of leaves}, $\mathit{leaves}(\tau)$, is the set of atoms
  occurring in the final goals of $D_{1},\ldots,D_{n}$.
\end{definition}

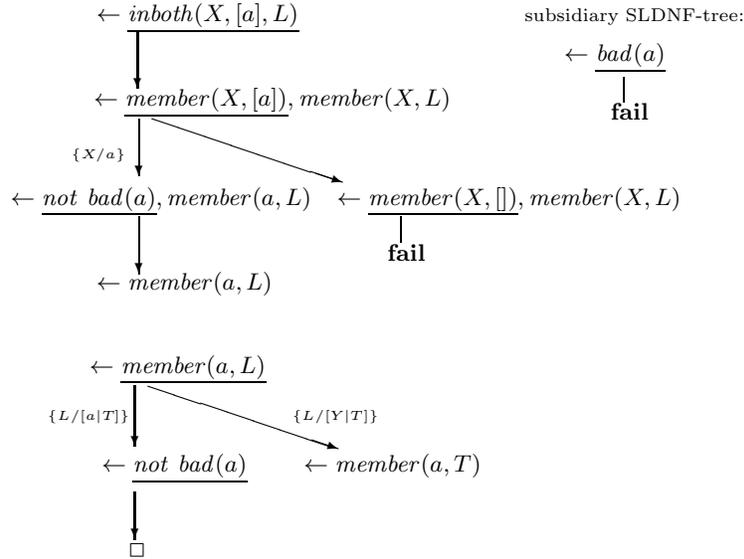
\begin{figure}[htbp] 
\setlength{\unitlength}{0.5pt} 
\begin{small}
\begin{picture}(455,212) 
\thinlines    
              \put(256,46){\line(0,-1){19}} 
              \put(246,13){{\bf fail}} 
              \put(68,120){\vector(3,-1){144}} 
              \put(208,55){$\leftarrow\underline{\mathit{member}(X,[])}, 
               \mathit{member}(X,L)$} 
              \put(58,46){\vector(0,-1){43}} 
              \put(-39,55){$\leftarrow \underline{\mathit{not\
                 bad}(a)}, \mathit{member}(a,L)$}
              \put(26,-9){$\leftarrow \mathit{member}(a,L)$} 
              \put(7,90){\tiny $\{X/a\}$} 
              \put(58,120){\vector(0,-1){43}} 
              \put(24,130){$\leftarrow\underline{\mathit{member}(X,[a])}, 
               \mathit{member}(X,L)$} 
              \put(26,194){$\leftarrow\underline{\mathit{inboth}(X,[a],L)}$}
              \put(57,185){\vector(0,-1){42}} 
              \put(350,194){{\scriptsize subsidiary SLDNF-tree:}} 
              \put(380,164){$\leftarrow \underline{\mathit{bad}(a)}$} 
              \put(425,152){\line(0,-1){19}} 
              \put(415,120 ){{\bf fail}} 
\end{picture}

\begin{picture}(455,200) 
\thinlines   
              \put(68,120){\vector(3,-1){144}} 
              \put(186,55){$\leftarrow \mathit{member}(a,T)$} 
              \put(33,55){$\leftarrow \underline{\mathit{not\
                  bad}(a)}$} 
              \put(58,40){\vector(0,-1){36}} 
              \put(53,-10){$\Box$} 
              \put(-8,94){\tiny $\{L/[a|T]\}$} 
              \put(177,94){\tiny $\{L/[Y|T]\}$} 
              \put(58,120){\vector(0,-1){46}} 
              \put(24,130){$\leftarrow\underline{\mathit{member}(a,L)}$} 
\end{picture}
\end{small} 
\caption{Incomplete SLDNF-trees for Example~\protect\ref{ex:inboth}}  
\protect\label{figure:inboth-ex}
\end{figure}

\begin{example} \label{ex:inboth} 
Let $P$ be the following program: 
\begin{small}\begin{pitemize} 
\item $\mathit{member}(X,[X|T])\leftarrow \mathit{not\ bad}(X)$ 
\item $\mathit{member}(X,[Y|T])\leftarrow \mathit{member}(X,T)$ 
\item $\mathit{inboth}(X,L1,L2)\leftarrow$ 
 $\mathit{member}(X,L1),\mathit{member}(X,L2)$ 
\item $\mathit{bad}(b)\leftarrow$ 
\end{pitemize}\end{small} 
Figure~\ref{figure:inboth-ex} represents an incomplete SLDNF-tree
$\tau$ for $P\cup\{\leftarrow \mathit{inboth}(X,[a],L)\}$.  This tree
has just one non-failing branch and the set of resultants
$resultants(\tau)$ contains the single clause:
\begin{quote} 
~~~~~$\mathit{inboth}(a,[a],L)\leftarrow \mathit{member}(a,L)$ 
\end{quote} 
Note that the complete SLDNF-tree for $P\cup\{\leftarrow
\mathit{inboth}(X,[a],L)\}$ is infinite.
\end{example}

With the initial goal atomic, the extracted resultants are program
clauses: the partial deduction of the atom.

\begin{definition}\label{def:partial-deduction} 
  Let $P$ be a normal program, $A$ an atom, and $\tau$ a finite
  non-trivial SLDNF-tree for $P \cup \{\leftarrow A\}$.
  Then the set of clauses $resultants(\tau)$ is called a {\bf partial
    deduction of $A$ in $P$}.
  If $\pdset{A}$ is a finite set of atoms, then a {\bf partial
    deduction of $\pdset{A}$ in $P$} is the union of the sets obtained
  by taking one partial deduction for each atom in $\pdset{A}$.
\end{definition}

In analogy with terminology in partial evaluation, the partial
deduction of $A$ in $P$ is also referred to as the {\em residual
  clauses\/} of $A$ and the partial deduction of $\pdset{A}$ in $P$ as
the {\em residual program\/}.

\begin{example} \label{ex:inboth-p2}
  Let us return to the program $P$ of Example~\ref{ex:inboth}.  Based
  on the trees in Figure~\ref{figure:inboth-ex}, we can construct the
  following partial deduction of $\pdset{A}$ =
  $\{\mathit{inboth}(X,[a],L), \mathit{member}(a,L)\}$ in $P$:

\begin{small}\begin{pitemize}
\item ~~~~~$\mathit{member}(a,[a|T])\leftarrow$
\item ~~~~~$\mathit{member}(a,[Y|T])\leftarrow \mathit{member}(a,T)$
\item ~~~~~$\mathit{inboth}(a,[a],L)\leftarrow \mathit{member}(a,L)$
\end{pitemize}\end{small}
\end{example}

Note that if $\tau$ is a trivial SLDNF-tree for $P\cup\{\leftarrow
A\}$ then $resultants(\tau) = \{A\leftarrow A\}$ and the specialised
program will be nonterminating for goals $\leftarrow A\theta$. The
problem is avoided by excluding trivial trees in
Definition~\ref{def:partial-deduction}.

The intuition underlying partial deduction is that a program $P$ can
be replaced by a partial deduction of $\pdset{A}$ in $P$ and that both
programs are equivalent with respect to queries which are constructed
from instances of atoms in $\pdset{A}$. A first issue to clarify is
what is intended by equivalent.  Lloyd and Shepherdson
\cite{Lloyd:jlp91} where the first to examine it in detail.  Using the
completion semantics as the declarative semantics, they can only show
soundness: that logical consequences from the completion of the
specialised program are also logical consequences of the completion of
the original program; the other direction, completeness (for instances
of atoms in $\pdset{A}$), does not hold in general, it holds only for
programs for which SLDNF is a complete proof procedure. Note that the
soundness result implies that answers obtained by SLDNF from the
specialised program are sound with respect to the original program for
any declarative semantics for which SLDNF is a sound procedure. For
procedural equivalence under the SLDNF proof procedure, Lloyd and
Shepherdson were able to obtain simple conditions guaranteeing
equivalence. The correctness with respect to the well-founded
semantics (now widely acknowledged to be better suited than completion
semantics to capture the meaning of logic programs
\cite{Denecker:TOCL01}) has been studied in
\cite{Seki:JLP93,Przymusinska:LPAR92,AravindanDung:NGC94}. The results
allow us to conclude that partial deduction, as defined above,
preserves declarative equivalence under the well-founded semantics for
ground atoms that are instances of $\pdset{A}$.  Almost all works on
partial deduction aim at preserving the procedural equivalence under
SLDNF. Before defining the extra conditions required to ensure it, we
introduce a few more concepts:

\begin{definition} \label{def:A-closed}
  \label{def:basics:independence}
  Let $A_1$, $A_2$, $A_3$ be three atoms, such that $A_3 =
  A_1\theta_1$ and $A_3 = A_2\theta_2$ for some substitutions
  $\theta_1$ and $\theta_2$.
  Then $A_3$ is called a {\bf common instance\/} of $A_1$ and $A_2$.
  Let $\pdset{A}$ be a finite set of atoms and $S$ a set containing
  atoms, conjunctions, and clauses.
  Then $S$ is {\bf $\pdset{A}$-closed\/} iff each atom in $S$ is an
  instance of an atom in $\pdset{A}$.
  Furthermore we say that $\pdset{A}$ is {\bf independent\/} iff no
  pair of atoms in $\pdset{A}$ has a common instance.
\end{definition}

The main result of \cite{Lloyd:jlp91} about procedural equivalence can
be formulated as follows:

\begin{theorem}[correctness of partial deduction]
 \label{theorem:pdcorrectness}
 
 Let $P$ be a normal program, $\pdset{A}$ a finite, independent set of
 atoms, and $P'$ a partial deduction of $\pdset{A}$ in $P$. For every
 goal $G$ such that $P'\cup \{ G \}$ is $\pdset{A}$-closed the
 following holds:

\begin{enumerate}
    \item $P'\cup \{G\}$ has an SLDNF-refutation with computed answer
    $\theta$ iff $P\cup \{G\}$ does.
    \item $P'\cup \{G\}$ has a finitely failed SLDNF-tree
    iff $P\cup \{G\}$ does.
\end{enumerate}
\end{theorem}

The theorem states that $P$ and $P'$ are procedurally equivalent with
respect to the existence of success-nodes and associated answers for
$\pdset{A}$-closed goals.  Furthermore, if we are in a setting where
SLDNF is complete for a particular declarative semantics then partial
deduction will preserve that semantics as well.  Among others, this is
the case for definite programs. For such programs the least Herbrand
models of $P$ and $P'$ will have the same intersection with the set of
$\pdset{A}$-closed ground atoms.  The fact that partial deduction
preserves equivalence only for $\pdset{A}$-closed goals distinguishes
it from e.g.\ unfold/fold program transformations which aim at
preserving equivalence for all goals.  Note that the theorem does not
tell us how to obtain $\pdset{A}$.  Also, it guarantees neither 
that termination, e.g.\ under Prolog execution, is preserved, nor that
computed answers are found in the same order.

Returning to Example~\ref{ex:inboth-p2}, we have that the partial
deduction of the set $\pdset{A}$ = $\{\mathit{inboth}(X,[a],L),$
$\mathit{member}(a,L)\}$ in $P$ satisfies the conditions of
Theorem~\ref{theorem:pdcorrectness} for the goals $\leftarrow
\mathit{inboth}(X,[a],[b,a])$ and $\leftarrow
\mathit{inboth}(X,[a],L)$ but not for the goal $\leftarrow
\mathit{inboth}(X,[b],[b,a])$. Indeed, the latter goal succeeds in the
original program but fails in the specialised one.
Intuitively, if $P'\cup\{G\}$ is not $\pdset{A}$-closed, then an
SLDNF-derivation of $P'\cup\{G\}$ may select a literal for which no
clauses exist in $P'$ while clauses did exist in $P$.  Hence, a query
may fail while it succeeds in the original program, or, due to
negation, may succeed while it fails in the original program.  If
$\pdset{A}$ is not independent then a selected atom may be resolved
with clauses originating from the partial deduction of two distinct
atoms. This may lead to computed answers that, although correct, are
not computed answers of the original program.  Moreover, this can in
turn lead to a specialised program that has a computed answer while
the original program flounders. The next example illustrates these
behaviours.

\begin{example}
  Take the following program $P$: 
  \begin{small}\begin{pitemize}
    \item $p(a,Y) \leftarrow q(Y)$
    \item $p(X,b) \larr$
    \item $q(c) \larr$
    \end{pitemize}\end{small}
  Let $\pdset{A} = \{ p(a,c) \}$.  A partial deduction $P'$ of
  $\pdset{A}$ in $P$ is:
  \begin{small}\begin{pitemize}
    \item $p(a,c) \larr$
    \end{pitemize}\end{small}
  $P'\cup\{\larr p(c,b)\}$ is not $\pdset{A}$-closed and
  $P'\cup\{\larr p(c,b)\}$ fails whereas $P\cup\{\larr p(c,b)\}$ does
  not.

  \noindent
  Now, let $\pdset{A}' = \{ p(a,X), p(Y,b) \}$.  A partial deduction
  $P''$ of $\pdset{A}'$ in $P$ is:
  \begin{small}\begin{pitemize}
    \item $p(a,c) \larr$
    \item $p(a,b) \larr$
    \item $p(X,b) \larr$
    \end{pitemize}\end{small}
  
  $\pdset{A}'$ is not independent and $P''\cup\{\larr p(Z,b)\}$
  produces the computed answers $\{Z/X\}$ and $\{Z/a\}$. The latter
  (redundant) answer is not produced by $P\cup\{\larr p(Z,b)\}$.
  Moreover, $P''\cup\{\larr p(Z,b),\neg p(a,Z)\}$ produces the
  computed answer $\{Z/a\}$ whereas $P\cup\{\larr p(Z,b),\neg
  p(a,Z)\}$ flounders. While one might consider this an improvement,
  it violates the requirement that the original and specialised
  program are procedurally equivalent for the goal.
\end{example}

Note that the original unspecialised program $P$ is also a partial
deduction of $\pdset{A}$ = $\{\mathit{member}(X,L),$
$\mathit{inboth}(X,L1,L2)\}$ in $P$, which furthermore satisfies the
correctness conditions of Theorem~\ref{theorem:pdcorrectness} for any
goal $G$.  In fact, one can always obtain the original program back by
putting into $\pdset{A}$ an atom $p(X_1,\ldots,X_n)$ for every
predicate symbol $p$ of arity $n$ and by constructing an SLDNF-tree of
depth 1 for every atom in $\pdset{A}$.  In other words, neither
Definition~\ref{def:partial-deduction} nor the conditions of
Theorem~\ref{theorem:pdcorrectness} ensure that any specialisation has
actually been performed.  Nor do they give any indication on how to
construct a suitable set $\pdset{A}$ and a suitable partial deduction
wrt $\pdset{A}$ satisfying the correctness criteria of the theorem.
These considerations are all generally delegated to the {\em
  control\/} of partial deduction, which we discuss in detail in the
following sections.

In the above development we deviated slightly from the original
presentation in \cite{Lloyd:jlp91}.  They define a {\em partial
  deduction of $P$ wrt $\pdset{A}$} to be ``a normal program obtained
from $P$ by replacing the set of clauses in $P$, whose head contains
one of the predicate symbols appearing in $\pdset{A}$ with a partial
deduction of $\pdset{A}$ in $P$.''  In other words, one keeps the
original definitions for those predicates which do not appear in
$\pdset{A}$.  Hence, Theorem~\ref{theorem:pdcorrectness} is a
corollary of the results in \cite{Lloyd:jlp91} and of the fact that
the original definitions are not reachable from any call which is
$\pdset{A}$-closed.  Note that our formulation, in contrast to
\cite{Lloyd:jlp91}, thus enables partial deduction to eliminate dead
code, i.e., code that can never be reached by executing a legal query
to the specialised program.  Hence, the original definition of
\cite{Lloyd:jlp91} is not used in any partial deduction (or even
partial evaluation) system we are aware of.

\begin{figure}[tbp]
\setlength{\unitlength}{0.55pt}
\begin{small}
        \centerline{
\begin{picture}(335,281)
\thinlines    \put(15,209){$\Box$}
              \put(139,13){$\larr \mathit{map}(\mathit{inv},T,Pt)$}
              \put(-29,13){$\larr \mathit{map}(\mathit{inv},T,Pt)$}
              \put(110,60){\vector(1,-1){34}}
              \put(100,60){\vector(-1,-1){32}}
              \put(62,254){\vector(-1,-1){32}}
              \put(103,70){$\larr \underline{\mathit{inv}(X,Px)},
\mathit{map}(\mathit{inv},T,Pt)$}
              \put(100,134){$\larr 
\underline{call(\mathit{inv}(X,Px))},
\mathit{map}(\mathit{inv},T,Pt)$}
              \put(104,119){\vector(0,-1){32}}
              \put(105,183){\vector(0,-1){32}}
              \put(101,195){$\larr
\underline{C=..[\mathit{inv},X,Px]},call(C), 
\mathit{map}(\mathit{inv},T,Pt)$}  
              \put(68,254){\vector(1,-1){34}}
              \put(10,263){$\larr
\underline{\mathit{map}(\mathit{inv},In,Out)}$}
              \put(-50,235){\tiny $\{In/[],Out/[]\}$}
              \put(103,235){\tiny $\{In/[X|T],Out/[Px|Pt]\}$}
              \put(114,165){\tiny $\{C/\mathit{inv}(X,Px)\}$}
              \put(137,40){\tiny $\{X/1,Px/0\}$}
              \put(-5,40){\tiny $\{X/0,Px/1\}$}\end{picture}}
\end{small}
        \caption{Unfolding Example~\protect\ref{ex:benefits-of-pd}}
        \protect\label{figure:benefits-of-pd}
\end{figure}
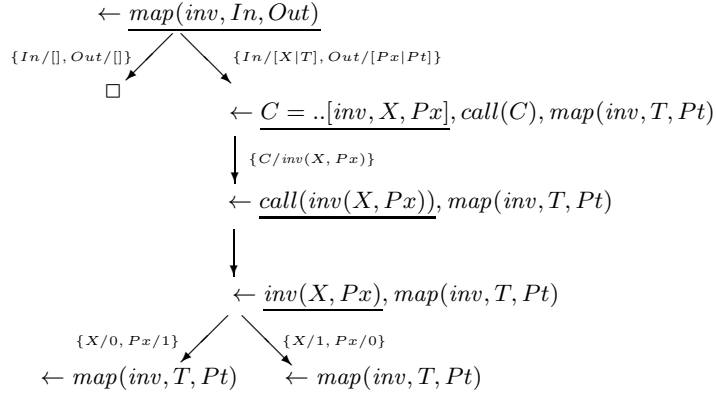

The following, more realistic example illustrates the practical
benefits of partial deduction.
\begin{example}
  \label{ex:benefits-of-pd}
  Let us examine the following program, defining the higher-order
  predicate $\mathit{map}$, which maps predicates over lists:

  \begin{small}\begin{pitemize}
    \item $\mathit{map}(P,[],[])\larr$
    \item $\mathit{map}(P,[X|T],[\mathit{Px}|\mathit{Pt}]) \larr$
      $C =.. [P,X,P\mathit{x}]$, $\mathit{call}(C)$,
      $\mathit{map}(P,T,P\mathit{t})$ 
    \item $\mathit{inv}(0,1) \leftarrow$
    \item $\mathit{inv}(1,0) \leftarrow$
    \end{pitemize}\end{small}
  
  Note that the above program can be seen as a pure definite logic
  program by conceptually adding a clause {\small
    $\mathit{call}(p(X_1,\ldots,X_n)) \leftarrow p(X_1,\ldots,X_n)$}
  for each n-ary predicate symbol $p$ and by adding a fact {\small
    $\mathit{{\small =..}}(f(X_1,\ldots,X_n), [f, X_1,\ldots,X_n])$}
  for each n-ary function symbol $f$.
  
  If we now want to map the $\mathit{inv}$ predicate on a list, then
  we can specialise the set $\pdset{A}$ =
  $\{ \mathit{map}(\mathit{inv},In,Out) \}$.  If we build the incomplete
  SLDNF-tree represented in Figure~\ref{figure:benefits-of-pd}, the set of
  all the
  leaf atoms is $\pdset{A}$-closed and we can construct the following residual
  program:

\begin{small}\begin{pitemize}
\item $\mathit{map}(\mathit{inv},[],[])\larr$
\item $\mathit{map}(\mathit{inv},[0|T],[1|\mathit{Pt}]) \larr$
 $\mathit{map}(\mathit{inv},T,\mathit{Pt})$
\item $\mathit{map}(\mathit{inv},[1|T],[0|\mathit{Pt}]) \larr$
 $\mathit{map}(\mathit{inv},T,\mathit{Pt})$
\end{pitemize}\end{small}

All the higher-order overhead (i.e., the use of $=..$ and
$\mathit{call}$) has been removed; also the calls to $\mathit{inv/2}$
have been unfolded.  When running the above programs on a set of
queries one notices that the specialised program runs up to 2 times
faster than the original one (depending on the particular Prolog
system used; and can be made even faster using filtering, as discussed
in Section~\ref{section:basics:renaming}).
\end{example}

The question that remains is, how do we come up with such (non-trivial
and correct) partial deductions in an automatic way?  This is exactly
the issue that is tackled in the remainder of this article.



\section{Main Control Issues}
\label{section:basics:control}

Partial deduction starts from an initial set of atoms $\pdset{A}$
provided by the user that is chosen in such a way that all runtime
queries of interest are $\pdset{A}$-closed. As we have seen,
constructing a specialised program requires to construct an SLDNF-tree
for each atom in $\pdset{A}$.  Moreover, one can easily imagine that
the conditions for correctness formulated in Theorem
\ref{theorem:pdcorrectness} may require to revise the set $\pdset{A}$.
Hence, when controlling partial deduction, it is natural to separate
the control into two components (as already pointed out in
\cite{Gallagher:PEPM93,MartensGallagher:ICLP95}):

\begin{itemize}
\item The {\em local control\/} controls the construction of the
  finite SLDNF-tree for each atom in $\pdset{A}$ and thus determines
  {\em what\/} the residual clauses for the atoms in $\pdset{A}$ are.
\item The {\em global control\/} controls the content of $\pdset{A}$,
  it decides {\em which\/} atoms are ultimately partially deduced
  (taking care that $\pdset{A}$ remains closed for the initial atoms
  provided by the user).
\end{itemize}

This gives rise to the generic scheme for a partial
deduction procedure (similar to the scheme in
\cite{Gallagher91:TR,Gallagher:PEPM93}) in
 Figure~\ref{figure:procedure1}.

\begin{figure}
\begin{procedurerm} 
  \label{algo:stdpd} 
  \noindent
  {\bf Input:} A program $P$ and a set $S$ of atoms of interest;  \\
  {\bf Output:} A specialised program $P'$ and a set of atoms $\pdset{A}$;

  \noindent
  {\bf Initialise:} $i = 0$, $\pdset{A}_0 =$ $S$;
 
  \noindent
  ~~{\bf repeat}
  \begin{zitemize}
  \item[] {\bf for} each $A_k \in \pdset{A}_i$ {\bf do}
  \item[~~~] {\bf let} $\tau_k:=\mathit{unfold}(P,A_k)$;
  \item[] {\bf let} $\pdset{A}'_i:= \pdset{A}_i \cup$
    $\{ B | B \in \mathit{leaves}(\tau_k) 
\}$;
  \item[] {\bf let} $\pdset{A}_{i+1}:= \mathit{revise}(\pdset{A}'_i)$; 
  \item[] {\bf let} $i:= i+1$
  \end{zitemize}

  \noindent
  ~~{\bf until} $\pdset{A}_{i} = \pdset{A}_{i-1}$;\\
   {\bf let} $\pdset{A}:=\pdset{A}_{i}$;\\
   {\bf let} $P' := \bigcup_{A_k\in \pdset{A}} \mathit{resultants}(\tau_k)$
\end{procedurerm}
        \caption{Generic partial deduction procedure}
        \label{figure:procedure1}
\end{figure}

The local control is exhibited by the function
$\mathit{unfold}(P,A_k)$ that returns a finite SLDNF-tree for $P \cup
\{\leftarrow A_k\}$.  Once all trees constructed, the atoms in their
leaves 
are added to the set of atoms. Then the global control, exhibited by
the function $\mathit{revise}(\pdset{A}'_i)$ is responsible for
adapting the set of atoms in such a way that all atoms
in $\pdset{A}'_i$ (and thus $S$  as well as all the leaves) 
 are
$\pdset{A}_{i+1}$-closed and that, eventually, a fixpoint is reached
where $\pdset{A}_{i} = \pdset{A}_{i-1}$ and a correct specialised
program can be extracted. The specialised program can then be used for
all queries that are $\pdset{A}$-closed.

To turn this scheme into a correct and usable algorithm, several
issues have to be considered. On the one hand, the specialised program
has to be correct and the partial deduction has to terminate.  On the
other hand, the specialised program should be as efficient as
feasible; it means that the available information, whether in the
input or in the context of calls to predicates, has to be exploited as
much as possible. These somewhat conflicting issues are elaborated
below:

\begin{enumerate}
\item {\em Correctness}. It requires that the specialised
  program computes the same results as the original for queries that
  are $\pdset{A}$-closed. Partial correctness is obtained by ensuring
  that Theorem~\ref{theorem:pdcorrectness} can be applied. This can be
  divided into a (very simple) local condition, requiring the
  construction of non-trivial trees, and into a global one related to
  the independence and closedness conditions. 
  
\item {\em Termination}.  There are two sources of potential
  nontermination.  First, one has to ensure that a finite SLDNF-tree
  is generated in finite time.  This is referred to as the {\em
    local\/} termination problem.  Secondly, one has to ensure that
  the iteration over the successive sets $\pdset{A}_i$ terminates and
  that the set itself remains finite (otherwise an infinite set of
  trees would have to be built).  This is referred to as the {\em
    global\/} termination problem. A related pragmatic aspect is that
  the partial deduction process finishes in a reasonable amount of
  time. What is reasonable depends on the application, e.g., whether the
  specialised program is to be used once or many times; whether the partial
  deduction process is part of standard compilation or a separate process
  initiated by the user.

\item {\em Degree of specialisation}.
  \label{pdaspects:precision}
  The degree to which the available information is exploited is called
  the {\em degree of specialisation\/} or {\em precision\/}, and
  unexploited information is referred to as {\em precision loss\/}. We
  can again discern two aspects.  One which we might call {\em local
    specialisation\/}. At first glance, the more atoms are unfolded,
  the more derivation steps are replaced by a single derivation step
  in the specialised program, hence the better the specialised program
  is.  However, as discussed in
  Section~\ref{subsection:efficiencydeterminacy}, one can unfold too
  much. Another issue related to local specialisation is that the
  atoms in a leaf of an SLDNF-tree are treated separately. No
  information is exchanged between the SLDNF-trees of distinct atoms.
  For instance, if we stop the unfolding process in
  Example~\ref{ex:inboth} for $G = \leftarrow
  \mathit{inboth}(X,[a,b,c],[c,d,e])$ at the goal $G' = \leftarrow
  \mathit{member}(X,[a,b,c]),$ $\mathit{member}(X,[c,d,e])$, partial
  deduction will not be able to infer the fact that the only possible
  answer for $G'$ and $G$ is $\{X/c\}$ as the atoms
  $\mathit{member}(X,[a,b,c])$ and $\mathit{member}(X,[c,d,e])$ are
  specialised separately.  (This problem is partially remedied by
  conjunctive partial deduction, c.f.\ Section~\ref{section:conjPD}.)
  Continuing the unfolding of $G' = \leftarrow
  \mathit{member}(X,[a,b,c]),$ $\mathit{member}(X,[c,d,e])$ achieves
  information propagation between the individual atoms and brings this
  fact to the surface, resulting in much better specialisation.
  
  The second aspect could be called the {\em global specialisation\/}
  and is related to the granularity of $\pdset{A}$.  In general having
  a more precise and fine grained set $\pdset{A}$ (with more {\em
    instantiated\/} atoms) will lead to better specialisation.  For
  instance, given the set $\pdset{A}$ = $\{ \mathit{member}(a,[a,b]),
  \mathit{member}(c,[d])\}$, partial deduction can perform much more
  specialisation (i.e., detecting that the goal $\leftarrow
  \mathit{member}(a,[a,b])$ always succeeds exactly once and that
  $\leftarrow \mathit{member}(c,[d])$ fails) than given the less
  instantiated set $\pdset{A}'$ = $\{ \mathit{member}(X,[Y|T])\}$,
  where $\mathit{member}(X,[Y|T])$ is the most specific atom which is
  more general than the atoms in $\pdset{A}$.

  A third aspect, orthogonal to both previous ones, is the size of the
  specialised program. Unfolding too much may result in code explosion,
  huge specialised programs, not only requiring lots of memory but
  perhaps also slowing down the execution. What counts for the user is
  not the amount of unfolding but the speed of the specialised
  program. Unfortunately, the actual performance is hard to predict and
  hence is not used to guide the specialisation process in current
  approaches. 

\end{enumerate}


\section{Local Control}
\label{section:basics:unfolding}

The function $\mathit{unfold}(P,A)$, introduced in the generic partial
deduction procedure of Section \ref{section:basics:control}, that
computes a finite SLDNF-tree for $P\cup\{\leftarrow A\}$ encapsulates
the local control and implements what is called an {\em unfolding
  strategy}. The unfolding strategy performs a finite number of
derivation steps, starting from the query $\leftarrow A$. It should
not be confused with the unfold rule in the unfold/fold program
transformation framework that performs a single derivation step on an
atom selected in a clause body.

The unfolding strategy applied on an atom $A$ determines exactly the
SLDNF-tree for that atom, hence its residual clauses. Consequently, it
has a big impact on the efficiency of the final program. In the next
section, we explain why too much unfolding can lead to inefficient
residual clauses and how such deterioration can be prevented.

\subsection{Efficiency by Determinacy}
\label{subsection:efficiencydeterminacy}

\begin{example}
  \label{app:toomuchunfolding}
  The well known append program is as follows:

  \begin{small}\begin{pitemize}
    \item $\mathit{app}([],L,L)\leftarrow$
    \item $\mathit{app}([H|X],Y,[H|Z])\leftarrow \mathit{app}(X,Y,Z)$
    \end{pitemize}\end{small}
  
  Now, let us try to specialise this program without having any
  partial input, i.e., $\pdset{A}$ = $\{\mathit{app}(X,Y,Z)\}$.  If we
  build an SLDNF-tree of depth 1 for $\mathit{app}(X,Y,Z)$ we just get
  the original program back. We have not obtained any improvements,
  but at least we have not worsened the program either. Actually,
  without any partial input, this is the best we can do.  Indeed, if
  we unfold more and, for example, perform two unfolding steps we
  obtain the following residual program:

  \begin{small}\begin{pitemize}
    \item $\mathit{app}([],L,L)\leftarrow$
    \item $\mathit{app}([X],L,[X|L])\leftarrow$
    \item $\mathit{app}([H,H'|X],Y,[H,H'|Z])\leftarrow \mathit{app}(X,Y,Z)$
    \end{pitemize}
  \end{small}
  
  Although the residual program performs only half of the resolution
  steps performed by the original program, it is not more efficient on
  standard Prolog implementations.  Indeed, the code size has
  increased and the resolution steps themselves have become more
  complicated.  Performing more unfolding steps makes things worse, as
  the following table shows (we ran a set of typical queries using
  SICStus Prolog 3.8.6 on a Linux'86 machine; relative runtimes are
  actual runtimes divided by runtime of the original program).

  \begin{center} 
    \begin{footnotesize}
      \begin{tabular}{l|llllllllllll}
        Unfolding Depth & 1 & 2 & 3 & 4 & 5 & 6 & 7 & 8 & 9 & 10 & 11 & 12\\ 
        Relative Runtime & 1 & 1.3 & 1.6 & 1.6 & 1.7 & 1.8 & 1.9 & 2.0
        & 2.0 & 2.2 & 2.4 & 2.5\\ 
      \end{tabular}
    \end{footnotesize}
  \end{center}

\end{example}

As the table shows, two extra unfolding steps already incur a
performance penalty of 60 \%.  This illustrates that too much
unfolding can seriously harm the efficiency of the residual program.
The result of such transformations may well be very implementation
dependent as not only unifications are more complex but also the
clause selection process. The overhead of the latter is dependent on
the quality of the indexing of the implementation. As the phenomenon
is typical for cases where the number of clauses increases, one could
call it {\em local code explosion\/} (there is a similar problem of
code explosion at the global level when the set $\pdset{A}$ gets too
large).

Another pitfall of too much unfolding is known as {\em work
  duplication\/}. The problem is illustrated in the following example.

\begin{figure}[htbp]
\begin{small}
        \centerline{\setlength{\unitlength}{0.55pt}
\begin{picture}(440,235)
\thinlines    \put(247,95){\vector(2,-1){60}}
              \put(303,13){{\bf fail}}
              \put(311,47){\line(0,-1){21}}
              \put(234,52){$\larr
\mathit{member}(a,L),\underline{\mathit{member}(a,[])}$}
\thicklines   \put(98,51){$\larr \mathit{member}(a,L)$}
\thinlines    \put(233,95){\vector(-2,-1){60}}
\thicklines   \put(155,103){$\larr
\mathit{member}(a,L),\underline{\mathit{member}(a,[Y])}$}
              \put(114,149){\vector(2,-1){60}}
              \put(14,103){$\larr \mathit{member}(a,L)$}
              \put(100,150){\vector(-1,-1){32}}
\thinlines    \put(11,164){$\larr
\mathit{member}(a,L),\underline{\mathit{member}(a,[X,Y])}$}
              \put(42,211){\vector(0,-1){30}}
              \put(10,217){$\larr\underline{\mathit{inboth}(a,L,[X,Y])}$}
              \put(45,137){\tiny $\{X/a\}$}
              \put(155,79){\tiny $\{Y/a\}$}
\end{picture}}
\end{small}
        \caption{Non-leftmost non-determinate unfolding for
        Example~\protect\ref{ex:inboth:duplication}}
        \protect\label{figure:inboth:duplication}
\end{figure}
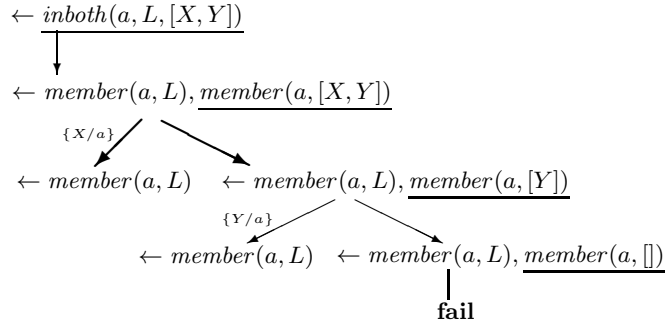

\begin{example}\label{ex:inboth:duplication}
Let $P$ be the following program (adapted from Example~\ref{ex:inboth}):

\begin{small}\begin{pitemize}
\item $\mathit{member}(X,[X|T])\leftarrow$
\item $\mathit{member}(X,[Y|T])\leftarrow \mathit{member}(X,T)$
\item $\mathit{inboth}(X,L1,L2)\leftarrow
\mathit{member}(X,L1),\mathit{member}(X,L2)$
\end{pitemize}\end{small}

Let $\pdset{A}$ = $\{\mathit{inboth}(a,L,[X,Y]),$
                       $\mathit{member}(a,L)\}$.
By performing the non-leftmost non-determinate unfolding
 for $\mathit{inboth}(a,L,[X,Y])$
 in Figure~\ref{figure:inboth:duplication}
 (and doing the same unfolding for
$\mathit{member}(a,L)$ as in Figure~\ref{figure:inboth-ex}), we obtain
 the following partial deduction $P'$ of $P$ with respect to $\pdset{A}$:

\begin{small}\begin{pitemize}
\item $\mathit{member}(a,[a|T])\leftarrow$
\item $\mathit{member}(a,[Y|T])\leftarrow \mathit{member}(a,T)$
\item $\mathit{inboth}(a,L,[a,Y])\leftarrow \mathit{member}(a,L)$
\item $\mathit{inboth}(a,L,[X,a])\leftarrow \mathit{member}(a,L)$
\end{pitemize}\end{small}

Let us examine the run-time goal
 $G = \leftarrow \mathit{inboth}(a,[h,g,f,e,d,c,b,a],[X,Y])$,
 for which $P'\cup\{G\}$ is $\pdset{A}$-closed.
Using the Prolog\index{Prolog} left-to-right computation rule the
 expensive sub-goal $\leftarrow \mathit{member}(a,[h,g,f,e,d,c,b,a])$
 is only evaluated once in the original program $P$,
 while it is executed twice in the specialised program $P'$.
 
 Observe that this is not a problem of local code explosion as in
 Example~\ref{app:toomuchunfolding}.  The increase from one to two
 $\mathit{inboth}/3$ clauses is arguably normal as calls to
 $\mathit{member}/2$ have been unfolded and this predicate is defined
 by two clauses.

\end{example}

Some partial evaluators, for instance, {\sc sage}
\cite{Gurr:LOPSTR93,Gurr:PHD} do not prevent such
work duplication.  This can result in arbitrarily big slowdowns, much
higher than those encountered in Example~\ref{app:toomuchunfolding}
(see, e.g., \cite{BowersGurr:metachapter}).

A common approach to prevent local code explosion and work duplication
relies on determinacy-based unfolding.  It was first proposed in
\cite{Gallagher91:ngc,Gallagher91:TR,Gallagher:PEPM93}.

\begin{definition}\label{def:determinateunfolding}
  The unfold function  is {\bf determinate\/} iff for every program
  $P$ and every goal $G$ it returns an SLDNF-tree with at most one
  non-failing branch.
\end{definition}

Applying determinate unfolding to an atom will produce an SLDNF-tree
with at most one resultant. Hence no local code explosion and no work
duplication can occur.  Also, determinacy is a strong indication that
enough input is available to select the ``right'' derivation, the
derivation that will be taken when the specialised program is executed
for the dynamic input.

Finally, determinate unfolding ensures that the order of
solutions, e.g., under Prolog execution, is not altered and that
termination is preserved (termination might however be improved, as
e.g., $\larr \mathit{loop},\mathit{fail}$ can be transformed into
$\larr \mathit{fail}$; for further details related to the preservation
of termination we refer to
 \cite{Proietti:PEPM91,BossiCocco:ALP94,BossiCoccoEtalle:LOPSTR95,%
LeuschelMartensSagonas:xsb}).

It is undecidable whether, for a given literal, one can construct
 an SLDNF-tree with at most one
  non-failing branch.
Hence, concrete unfold functions use a so-called {\em lookahead\/}
 to decide whether a particular literal can be unfolded.
Using a lookahead of 0 means that a literal can only be unfolded
 if it produces one resultant or less, while using a lookahead of 1 means
 that we can also select literals which produce more than one resultant,
 provided that all but one of them fail at the next resolution step.

\begin{figure}[htbp]
\begin{center}
\setlength{\unitlength}{0.0003in}%
\begin{picture}(8106,2406)(514,-1723)
\thicklines
\put(1126,389){\vector(-1,-1){600}}
\put(1276,389){\vector( 1,-1){600}}
\put(1201,389){\vector( 0,-1){600}}
\put(1201,-361){\vector( 0,-1){600}}
\put(1876,-361){\vector( 0,-1){600}}
\put(526,-361){\vector( 0,-1){600}}
\put(526,-1111){\vector( 0,-1){600}}
\put(1876,-1111){\vector( 0,-1){600}}
\put(3601,389){\vector( 0,-1){600}}
\put(3451,-361){\vector(-1,-1){600}}
\put(3601,-361){\vector( 0,-1){600}}
\put(3751,-361){\vector( 1,-1){600}}
\put(2851,-1111){\vector( 0,-1){600}}
\put(4351,-1111){\vector( 0,-1){600}}
\put(6001,389){\vector( 0,-1){600}}
\put(6001,-361){\vector( 0,-1){600}}
\put(6001,-1111){\vector( 0,-1){600}}
\put(6151,-1111){\vector( 1,-1){600}}
\put(5851,-1111){\vector(-1,-1){600}}
\put(7951,389){\vector( 0,-1){600}}
\put(7951,-361){\vector( 0,-1){600}}
\put(7951,-1111){\vector( 0,-1){600}}
\put(1276,539){\makebox(0,0)[b]{\small shower}}
\put(3601,539){\makebox(0,0)[b]{\small fork}}
\put(6001,539){\makebox(0,0)[b]{\small beam}}
\put(7951,539){\makebox(0,0)[b]{\small pure}}
\end{picture}
\end{center}
\caption{Three almost determinate trees
 and one determinate tree}\protect\label{figure:det-unfolding-rules}
\end{figure}
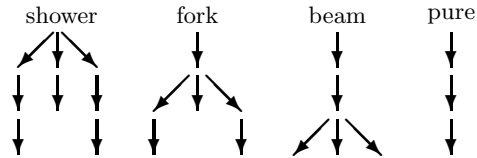

The determinate unfolding approach is too restrictive, as  we have to
prevent trivial trees, and is usually replaced by {\em almost determinate
  unfolding\/} that allows one non-determinate unfolding step. This
non-determinate step may either occur only at the root (used, e.g., in
\cite{Gallagher91:TR}), only at the bottom (used in
\cite{Gallagher91:ngc,LeuschelDeSchreye97:ngc}), or anywhere in the
tree (an option which can be used within {\sc ecce}
\cite{Leuschel96:ecce-dppd}).  These three forms of {\em almost
  determinate\/} trees are illustrated in
Figure~\ref{figure:det-unfolding-rules}.
However, as the experiments
 in \cite{LeuschelMartensDeSchreye:Toplas} show,
 even almost determinate unfolding can be too restrictive and
 does not fare very well on highly non-deterministic programs,
 such as the ``contains'' benchmark
 \cite{Leuschel96:ecce-dppd}
 devised by Lam and Kusalik.
Nonetheless, as we will see in Section~\ref{section:conjPD},
  this is much less of an issue in the setting of so-called
  conjunctive partial deduction.

To avoid the work duplication pitfall described in
Example~\ref{ex:inboth:duplication}, the one non-determinate unfolding
step performed by an almost determinate unfolding rule should mimic
the runtime selection rule (leftmost for Prolog).
Observe that for a shower tree
this is always satisfied, as there is only one literal in the
root.

Among the three almost determinate unfolding trees, the shower
 is the most restrictive one as it only allows a
 non-determinate step if necessary to avoid a trivial tree.
All three avoid local code explosion as the number of residual
clauses cannot exceed the number of program clauses defining the atom
selected at the non-deterministic step.

Unfortunately, fork and beam determinate unfolding 
can still lead to duplication of work, namely in unification with
multiple heads:

\begin{example}\label{ex:inboth:duplication2}
  Let us adapt Example~\ref {ex:inboth:duplication} by using
  $\pdset{A}$ = $\{\mathit{inboth}(X,[Y],[V,W])\}$.  We can fully
  unfold $\leftarrow \mathit{inboth}(X,[Y],[V,W])$ and we then obtain
  the following partial deduction $P'$ of $P$ with respect to
  $\pdset{A}$:

  \begin{small}\begin{pitemize} 
    \item $\mathit{inboth}(X,[X],[X,W])\leftarrow$
    \item $\mathit{inboth}(X,[X],[V,X])\leftarrow$
    \end{pitemize}\end{small}

\end{example}

No goal has been duplicated by the leftmost non-determinate unfolding,
but the unification $X\!=\!Y$ for $\leftarrow
\mathit{inboth}(X,[Y],[V,W])$ has been duplicated in the residual
code.  This unification can have a substantial cost when the
corresponding actual terms are large. In fact, code like the above
could as well be written by hand, and the problem could be attributed
to poor compiler technology. We are here touching upon a rather low
level issue on the borderline between specialisation and compilation
that is not well mastered and not much studied. Ideally, unfolding
decisions should be based on a more precise performance model that
takes into account the compiler technology of the target system such
as clause indexing, the cost of term construction operations, and the
cost of having too many arguments (often considerable slowdown occurs
if the number of arguments exceed 32).  In the absence of such
detailed modelling and of better compiler technology, pragmatic
solutions are either to use shower determinate unfolding only, or to
provide a postprocessor that avoids the unification overhead through
the introduction of explicit disjunctions (denoted ``;'' as in
Prolog):

\begin{small}\begin{pitemize}
\item $\mathit{inboth}(X,[X],[V,W])\leftarrow
(X\!=\!V)
~;~
(X\!=\!W)$
\end{pitemize}\end{small}

\noindent
or, even better on most Prolog systems\footnote{Private communication
  from Bart Demoen.}, through the introduction of an auxiliary
predicate (so called transformational indexing):

\begin{small}\begin{pitemize}
\item $\mathit{inboth}(X,[X],[V,W])\leftarrow \mathit{one\_of}(X,V,W)$
\item $\mathit{one\_of}(X,X,\_) \leftarrow$
\item $\mathit{one\_of}(X,\_,X) \leftarrow$
\end{pitemize}\end{small}


\subsection{Ensuring Termination}

Having solved the problems of local code explosion and work
duplication, we still have no adequate unfolding function. Indeed
almost determinate unfolding can result in infinite branches.  In
(strict) functional programs such a condition is equivalent to an
error in the original program.  In logic programming (and in lazy
functional programming) the situation is somewhat different: a goal
can infinitely fail (in a deterministic way) during partial deduction
but still finitely fail at run time, i.e., when executed using fully
instantiated input.  In applications where one searches an infinite
space for the existence of a solution (e.g.\ theorem proving) even
infinite failures (i.e., infinite SLDNF-trees without a refutation in
the main tree) at run-time do not necessarily indicate an error in the
program: they might simply be due to non-existence of a solution.
This is why, perhaps in contrast with functional programming,
additional measures on top of determinacy should be adopted to ensure
local termination.
 
Early approaches either did not guarantee termination or made ad-hoc
decisions to enforce termination. Subsumption checking (unfolding
stops when the selected atom is an instance of a previously selected
atom) and variant checking (unfolding stops when the selected atom is
a variant of a previously selected atom) are examples of the former
approach and are mentioned in
\cite{TakeuchiFurukawa:ifip86,FullerAbramsky:88,%
  LeviSardu:ngc88,Benkerimi:naclp90,vanHarmelen:Book89} but are
inadequate \cite{Bruynooghe:ngc92} as the following examples
illustrate. 

\begin{example} \label{ex:revacc}
Take the following simple program for reversing a list.

\begin{small}\begin{pitemize}
\item $\mathit{rev}([],\mathit{Acc},\mathit{Acc})\leftarrow$
\item $\mathit{rev}([H|T],\mathit{Acc},\mathit{Res})\leftarrow$
         $\mathit{rev}(T,[H|\mathit{Acc}],\mathit{Res})$
\end{pitemize}\end{small}

Unfolding $\larr rev(X,[],R)$ using subsumption or variant
 checking will give rise to an infinite SLD-tree.
\end{example}

The use of an arbitrary depth bound is an example of an ad-hoc
approach. Unavoidably, there are cases where this leads to either too much
unfolding and code explosion, or too little unfolding and under
utilisation of the available information. The hope is that the other
components of the unfolding strategy will cause that the depth bound
is used only in pathological cases. Approaches using depth
bounds are in 
\cite{Venken:ECAI84,Prestwich:PEPM93,FullerBocicBertossi:NGC96,%
Sahlin93:ngc,Sahlin:phd}.


\subsubsection{Offline approaches}
\label{sec:offline}

One approach to ensure termination is to perform a preliminary
analysis and to use the results of this analysis to make the unfolding
decisions.

\paragraph{{\bf 1.} Offline Annotations.} \label{subsection:offline}
In this approach, often referred to as {\em offline\/} (because almost
all the control decisions are taken before the actual specialisation
phase), unfolding proceeds in a strict left-to-right fashion and every
call in the program to be specialised has an annotation specifying
whether it is to be unfolded or not. In the latter case the call is
said to be residualised. One could annotate the programs by hand and
then check whether the annotation is correct, i.e.\ the unfolding will
terminate. This can be achieved by removing the literals annotated as
to be residualised (as they are residualised, they are not executed
and do not create bindings) and to use existing tools for termination
analysis of logic programs (see \cite{DeSchreyeDecorte:JLP94} for a
survey and the specialised literature for more recent work).
It is a component of the approach of \cite{Vanhoof:LPAR01} described
at the end of the next paragraph.

However, in general one also wants to automatically derive the
annotations itself: this preliminary analysis is referred to as a {\em
  binding-time analysis\/} (bta).
The first fully implemented bta for logic programs was probably
presented in \cite{Gurr:PHD}, for the {\sc sage} system.  This bta is
monovariant and unfolding decisions are taken at the predicate
 level, i.e., for each predicate all calls are
  either unfolded or residualised.
This is thus still too restrictive in practice. 
A more recent and more powerful bta (for functional programs), which ensures
 termination and can even handle sophisticated programs
 such as interpreters, is
presented in \cite{GlenstrupJones:96:BTATermination}.
\cite{BruynoogheLeuschelSagonas:ESOP98} presented a step towards a
polyvariant bta for logic programs.
Assuming an unfolding condition for every
predicate is given, it employs abstract interpretation to derive a polyvariant
version of the original program where every call is annotated with an
unfolding decision (for some predicates, the clauses defining them can
be multiplied and each version is differently annotated).
\cite{VanhoofBruynooghe:ICLP99} have developed a binding time analysis
for Mercury \cite{SomogyiHendersonConway:jlp}, a typed and moded logic
programming language. Given the features of Mercury, this work is
closer to work in partial evaluation of functional programs than to
partial deduction of logic programs.
\cite{Vanhoof:LPAR2000} has extended it to cope with the higher-order
features and module structure of Mercury.
Finally, \cite{Vanhoof:LPAR01} describes a full binding time analysis
for logic programs. The termination analyser of \cite{codish99:jlpb}
has been extended for the case that it cannot prove termination. The
extension identifies the atoms in clause bodies that are at the origin
of the failure to prove termination. This termination analyser is then
used in an iterative process. When it proves termination, all calls
are annotated as unfoldable. In the other case, one of the identified
atoms is annotated as to be residualised and the program with the
residualised atom removed is again analysed for termination.
Eventually, enough atoms are annotated as residualised to allow a
proof that the execution (unfolding) terminates.
  
One of the big advantages of the offline approach is the efficiency of
the specialisation process itself: indeed, once the annotations have
actually been derived (automatically by the above btas or by hand),
the specialiser is relatively simple, and can be made to be very
efficient, since all
decisions concerning local control are made before and not
during specialisation.

The simplicity of the specialiser also means that it is much easier to achieve 
 {\em self-application}, i.e., specialise the specialiser
 itself using partial evaluation.
Indeed,  achieving effective self-application
 was one of the initial motivations for investigating offline control techniques
\cite{JonesSestoftSondergaard:LSC89}.
Self-application was first achieved in the logic programming context
 in \cite{MogensenBondorf:LOPSTR92} for a subset of Prolog
 and later in \cite{Gurr:LOPSTR93,Gurr:PHD} for full G\"{o}del.
Self-application enables a partial evaluator
 to generate so-called ``compilers''
 from interpreters using the second Futamura projection
 and a compiler generator ($\mathit{cogen}$) using
  the third Futamura projection  (see, e.g., \cite{Jones:peval}).
However, the actual creation of the $\mathit{cogen}$
 according to the third Futamura projection
 is not of much interest to users since $\mathit{cogen}$
 can be generated once and for
 all when a specialiser is given.
This is known as the {\em cogen-approach\/}
 and has been successfully applied in many programming
 paradigms
 \cite{Beckman:76,Romanenko:88,Holst:89a,HolstLaunchbury:92,%
BirkedalWelinder:94:handcogen,Andersen94:PhD}.
In the logic programming setting,
\cite{Neumann:META90,Neumann:LPAR90} presents
 a system for definite clause grammars
 which is very similar to a $\mathit{cogen}$,
 but not from a partial evaluation perspective.
The first $\mathit{cogen}$ for a logic programming
 language was thus (arguably)
presented in
\cite{JorgensenLeuschel:Cogen,LeuschelJorgensen:WOID99}.
The resulting system {\sc logen} performs the unfolding at
speeds similar to
ordinary execution, and is thus well suited for
 applications, 
 where speed of the specialisation is crucial (and where
 the program to be specialised can be analysed beforehand by the
 bta).

\paragraph{{\bf 2.} Delay declarations.}

Instead of taking all unfolding decisions at analysis time, one can
also infer conditions under which unfolding is guaranteed to terminate
and leave it to the specialiser to check whether a particular atom
meets the condition and can be unfolded. The specialiser, knowing the
actual static input, may then be able to unfold more atoms than a
binding time analyser would consider safe.
The required analysis has lots in common with the analysis used for
logic programs with {\em delay declarations\/} (also called
coroutining). When executing such programs, calls are suspended until
they meet their delay declarations. Analysis can be developed that can
verify whether the program terminates for a given delay declaration or
that can infer delay declarations ensuring termination.
Relevant work is in
\cite{naish:ctl:92,Luttringhaus93,Marchiori:ILPS95,MartinK97}.  Using
the delay declarations for which the program terminates to decide
whether atoms should be unfolded or residualised ensures termination
of unfolding (Incomplete branches in the SLDNF-tree correspond to
deadlocked derivations).

Such an approach has actually not been very widely used yet, with the
exception of \cite{FujitaFurukawa88:ngc},
\cite{Leuschel:LOPSTR94,LeuschelDeSchreye98:jlp} and
\cite{MartinLeuschel:PSI99,Martin:phd}.  Note that some of the delay
declarations derived by \cite{naish:ctl:92,Marchiori:ILPS95,MartinK97}
can be overly restrictive in the context of unbounded (i.e., partially
instantiated) datastructures (common in partial deduction).  Hence,
\cite{MartinLeuschel:PSI99,Martin:phd} extend this approach by
pre-computing minimum sizes for the unbounded structures and unfold
atoms as long as sizes remain under the minimum.


\subsubsection{Online Approaches: Well-founded and Well-quasi orders}

In this section we look at so called online approaches that monitor
the growth of branches of SLDNF-trees, continue unfolding as long as
there is some evidence that interesting computations are performed but
are also guaranteed to terminate. To achieve this, they maintain orders
over the nodes of a branch that are chosen in such a way that infinite
branches are impossible. If care is taken that there cannot be an
infinite number of attempts to rebuild a branch, the construction of
the tree must terminate.

Well-founded orders and well-quasi orders are well known to allow the
definition of admissible sequences that are always finite. Their
definitions are as follows:

\begin{definition}\label{def:po}\label{def:nonstrictpo}
  A {\bf strict partial order\/}  $<_S$ 
on a set $S$ is an irreflexive, transitive, and thus
  asymmetric binary relation on $S$.
  A {\bf quasi order\/}\index{quasi order} (also called preorder)
  $\leq_S$ on a set $S$ is a reflexive and transitive binary relation
  on $S$.
\end{definition}

\begin{definition}\label{def:wfoset}\label{def:wfo}
  Let $<_S$ be a strict partial order on a set $S$.
  A sequence of elements $s_1,s_2,\ldots$ in $S$ is called {\bf
    admissible with respect to $<_S$\/} iff $s_i+1<s_{i}$, for all
  $i\geq 1$.
  The relation $<_S$ is a {\bf well-founded order (wfo)\/} iff there
  is no infinite admissible sequence with respect to $<_S$.
\end{definition}

\begin{definition}\label{def:wqo}\label{defwqoset}
  Let ${\leq}_S$ be a binary relation on $S$. 
  A sequence of elements $s_1,s_2,\ldots$ in $S$ is called {\bf
    admissible with respect to ${\leq}_S$\/} iff there are no $i < j$
  such that $s_i~{\leq}_S~s_j$.
  The relation ${\leq}_S$ is a {\bf well-binary relation (wbr)\/} on
  $S$ iff there are no infinite admissible sequences with respect to
  ${\leq}_S$.
  The relation ${\leq}_S$ is a {\bf well-quasi order (wqo)\/} on $S$
  iff it is a well-binary relation and a quasi order.
\end{definition}

In what follows, 
we define an {\em expression\/} to be either a term, an atom, a
conjunction, or a goal.

When defining orders over the sequence of
 nodes in a branch, nobody has found it
useful to compare complete goals, only the selected atoms are
compared. Also, it was quickly realised that it was difficult to
define an order relation on the full sequence that was giving good
unfoldings and that it was sufficient and easier to do so on certain
subsequences. The essence of the most advanced technique, based on
covering ancestors \cite{Bruynooghe:ngc92} can be captured in the
following definitions.

\begin{definition}
  If a program clause $H \leftarrow B_1, \ldots, B_n$ is used in a
  derivation step with selected atom $A$ then, for each $i$, $A$ is
  the {\bf parent\/} of the instance of $B_i$ in the resolvent and in
  each subsequent goal where an instance originating from $B_i$
  appears (up to and including the goal where $B_i$ is selected).
  The {\bf ancestor\/} relation is the transitive closure of the parent
  relation.
\end{definition}

\begin{definition}
  Let $G_0$, $G_1$, \ldots, $G_n$ be an SLDNF-derivation with selected
  atoms $A_1$,$A_2$, \ldots, $A_{n}$. 
  
  The {\bf covering ancestor sequence of $A_i$}, a selected atom, is
  the maximal subsequence $A_{j_1}$, $A_{j_2}$, \ldots $A_{j_m} =
  A_i$ of $A_1$, $A_2$, \ldots, $A_{i}$ such that all atoms in the
  sequence have the same predicate symbol and, for all $1 \leq k < m$
  it holds that $A_{j_k}$ is an ancestor of $A_{j_{k+1}}$.
  
  An SLDNF-derivation $G_0$, $G_1$, \ldots, $G_n$ is {\bf safe with
  respect to an order\/} (wfo or wqo) if all covering ancestor sequences
  of the selected atoms are admissible with respect to that order.
  
\end{definition}

Covering ancestors, first introduced for well-founded orders
\cite{Bruynooghe:ngc92} and later also used with well-quasi orders
(e.g., \cite{LeuschelMartensDeSchreye:Toplas}), are so useful because an
infinite derivation must have at least one infinite covering ancestor sequence.
Hence, an atom can be unfolded when the SLDNF-derivation remains safe.
Moreover, experience has shown that the admissibility of the covering
ancestor sequences is a strong indication that some interesting
specialisation is going on.

\paragraph{{\bf  Well-founded orders.}}

Inspired by their usefulness in the context of static termination
analysis (see e.g.,
\cite{DershowitzManna:CACM79,DeSchreyeDecorte:JLP94}), well-founded
orders have been successfully employed to ensure termination of
partial deduction in
\cite{Bruynooghe:ngc92,MartensDeSchreyeHorvath:TCS94,%
  MartensDeSchreye:jlp95b,Martens:phd}.  In addition, the unfolding
performed by these techniques is related to the structural aspect of
the program and goal to be partially deduced.  They are arguably the
first theoretically and practically satisfying solutions for the local
termination problem.

\begin{example}
  A simple well-founded order can be obtained by comparing the {\em
    termsize\/} of atoms: we say that $A<B$ iff
  $\mathit{termsize}(A)<\mathit{termsize}(B)$, where
  $\mathit{termsize}(t)$ of an expression $t$ is the number of
  function and constant symbols in $t$.  Let us apply this to the
  $\mathit{member}$ program $P$ of Example~\ref{ex:inboth}.  Based on
  that wfo, the SLDNF-tree with successive goals $\leftarrow
  \mathit{member}(X,[a,b|T])$, $\leftarrow \mathit{member}(X,[b|T])$
  and $\leftarrow \mathit{member}(X,T)$ results in the covering
  ancestor sequence $\mathit{member}(X,[a,b|T])$,
  $\mathit{member}(X,[b|T])$, $\mathit{member}(X,T)$ which is
  admissible because the termsize of the selected atoms strictly
  decreases at each step.  However, it is not allowed to perform a
  further unfolding step as the addition of the element
  $\mathit{member}(X,T')$ to the covering ancestor sequence makes the
  sequence inadmissible.
\end{example}

In general, measuring just the termsize of atoms leads to overly
conservative unfolding.  Take for example the $\mathit{rev}$ program
from Example~\ref{ex:revacc}.  Given, e.g., the goal $\larr
rev([a,b],[],R)$ one would ideally want to achieve full unfolding.
Fully unfolding $\larr rev([a,b],[],R)$ results in a covering ancestor
sequence $rev([a,b],[],R)$, $rev([b],[a],R)$, $rev([],[b,a],R)$.
Unfortunately, as the termsize is 6 for all the elements, the sequence
is not admissible and the derivation is not safe.  However, using a
wfo which just examines the termsize of the first argument, the branch
is admissible and full unfolding can be achieved. This illustrates
that it is difficult to decide beforehand which is the wfo that gives
the best unfolding and that there is a need to adjust the wfo while
unfolding.

Such an approach is followed in 
\cite{Bruynooghe:ngc92,MartensDeSchreyeHorvath:TCS94,%
MartensDeSchreye:jlp95b,Martens:phd}.  They start off with a simple
wfo and then refine it during the unfolding process.

\begin{example} \label{example:refine-wfo}

  Consider a query $G_{1}$ = $\larr \mathit{rev}([a,b|T],[],R)$ for
  the $\mathit{rev}$ program $P$ of Example~\ref{ex:revacc}. One
  starts with the wfo based on summing up the termsizes of the
  arguments whose positions are in the set $S_1=\{1,2,3\}$.  Unfolding
  one step, the resolvent is $G_{2}$ = $\larr
  \mathit{rev}([b|T],[a],R)$ and the covering ancestor sequence is
  $\mathit{rev}([a,b|T],[],R)$, $\mathit{rev}([b|T],[a],R)$. Using the
  wfo based on $S_1$, both atoms have size 5 and the covering ancestor
  sequence is inadmissible. The adjustment of the wfo removes a
  minimal number of elements from $S_1$ such that the sequence becomes
  admissible. Using $S_2 = \{1,3\}$ achieves this. Another unfolding
  step yields the goal $G_{3}$ = $\larr \mathit{rev}(T,[b,a],R)$ and
  the covering ancestor sequence remains admissible. Performing
  another unfolding step results in the goal $\larr
  \mathit{rev}(T',[H',b,a],R)$ and the covering ancestor sequence
  $\mathit{rev}([a,b|T],[],R)$, $\mathit{rev}([b|T],[a],R)$,
  $\mathit{rev}(T,[b,a],R)$, $\mathit{rev}(T',[H',b,a],R)$, which is
  {\em not\/} admissible for $S_2$ and for any subset of it. Hence it
  is not allowed to perform the last step.

\end{example}

The above example suggests two critical points.  First, one has to
ensure that one cannot continuously refine a wfo. In the above example
this was ensured by only allowing arguments to be removed.  In a more
general setting (e.g., where one can vary weights associated with
constants and function symbols) one has to ensure that the successive
wfos are themselves well-founded.

Secondly, when selecting a new wfo, verifying that the last atom in
the covering ancestor sequence is strictly smaller than the previous
one does not guarantee that the whole sequence is admissible (while it
suffices when extending an admissible sequence for a given wfo with
one atom).  Hence, early algorithms tested the whole sequence on
admissibility.  This can be expensive for long sequences.

\cite{MartensDeSchreye:jlp95b,Martens:phd} therefore advocates another
 solution: not re-checking the entire sequence on the grounds
 that it does not threaten termination (provided that the refinements
 of the wfo
  themselves are well-founded).
This leads to sequences $s_{1},s_{2},\ldots$ of selected
 literals which are not well-founded
 but {\em nearly-founded\/} \cite{MartensDeSchreye:jlp95b,Martens:phd}
 meaning that
 $s_{i} \not < s_{j}$ only for a finite number of pairs
 $(i,j)$ with $i>j$.
This improves the efficiency of the unfolding process, but
 has the tradeoff that it
  can lead to
 sequences of covering ancestors
 which contain more than one occurrence of exactly the same
 selected literal \cite{Leuschel:LOPSTR98},
 which is considered a clear sign of too much
 unfolding. 

\paragraph{{\bf Well-quasi orders.}}

A drawback of the above mentioned wfo approaches, is that they will
not be able to satisfactorily handle certain programs.
For example, Datalog programs (logic programs without
 functors) will pose problems
 as all constants have the same
 size under the measures that are typically used in wfos.
Assigning a different size to each constant will not solve the problem.
As the ordering is total, there will be situations where it
leads to suboptimal unfolding.
For Datalog program on could use
variant checking as the number of distinct variants is finite. A more
fundamental solution is to use quasi orderings.

Local termination is ensured in a similar manner as for wfos by
allowing only safe SLDNF-trees.
The difference is that the admissibility of  covering ancestor sequences 
is based on well-quasi orders.
Hence an element added to an admissible sequence is not necessarily
strictly smaller than all elements in the sequence as is the case for
a wfo. 
This, e.g., allows a wqo to have no a priori fixed size or order
attached to functors and arguments
and avoids to focus in advance on specific
sub-terms. The latter is  
crucial to obtain good unfolding
of metainterpreters \cite{Leuschel:SAS98,Leuschel:LOPSTR98}.

The first explicit uses of wqos to ensure termination of partial
deduction are in \cite{bol:jlp93-pd,Sahlin93:ngc}.
\cite{Prestwich92:TR} presents a method which can be seen as a simple
wqo: it maps atoms to so-called ``patterns'' (of which there are only
finitely many) and unfolds every pattern at most once.
\cite{Prestwich92:TR} also presents an improvement whereby it is
always allowed to decrease the {\em termsize\/}. This can still be
seen as a wqo.  In fact,
every wfo can be mimicked by a wqo
and the combination of two wqos is still a wqo
 \cite{Leuschel:SAS98,Leuschel:LOPSTR98}.

An interesting wqo is the homeomorphic embedding relation $\homeo$,
which derives from results by \cite{Higman:PLMS52} and
\cite{Kruskal:TAMS60}.  It has been used in the context of term
rewriting systems in \cite{Dershowitz:jsc87,Dershowitz:htcs90}, and
adapted for use in supercompilation in \cite{SorensenGluck:ILPS95}.

What follows is an adaptation of the
 definition from \cite{SorensenGluck:ILPS95}, in turn based
on the so-called pure $\homeo$ in \cite{Dershowitz:htcs90}. It
has a simple treatment of variables.

\begin{definition} \label{definition:homem}
\label{def:homeo}
The {\bf  homeomorphic embedding\/} relation
  $\homeo$ on terms and atoms
 is defined inductively as follows (i.e.\ $\homeo$
 is the least relation satisfying the rules), where $n\geq 0$,
 $p$ denotes predicate symbols,
 $f$ denotes function symbols, and $s,s_1,\ldots,s_n,t,t_1,\ldots,t_n$
 denote terms:
\begin{pitemize}
\item[1.] $X \homeo Y$ for all variables $X,Y$
\item[2.] $s \homeo f(t_{1},\ldots,t_{n})$
  if $s \homeo t_{i}$ for some $i$
\item[3.] $f(s_{1},\ldots,s_{n}) \homeo f(t_{1},\ldots,t_{n})$ if
        $\forall i\in \{1,\ldots,n\}:$ $s_{i} \homeo t_{i}$.
\item[4.] $p(s_{1},\ldots,s_{n}) \homeo p(t_{1},\ldots,t_{n})$ if
        $\forall i\in \{1,\ldots,n\}:$ $s_{i} \homeo t_{i}$.
\end{pitemize}
\end{definition}

When $s\homeo t$ we also say that $s$ is {\em embedded in} $t$ or $t$
is {\em embedding} $s$.  By $s\homeostrict t$ we denote that $s\homeo
t$ and $t\nothomeo\, s$.  The important property is that $\homeo$ is a
well-quasi order \cite{SorensenGluck:ILPS95}.

The intuition behind the above definition is that $A\homeo B$ iff $A$
can be obtained from $B$ by removing some symbols i.e.\ that the
structure of $A$, splitted in parts, reappears within $B$.  For
instance we have $p(a) \homeo p(f(a))$ because $p(a)$ can be obtained
from $p(f(a))$ by removal of ``$f()$'' Observe that the removal
corresponds to the application of rule 2 \condtext{}{(cf.\ 
  Lemma~\ref{lemma:diving-eq})} (also called the diving rule) and that
we also have $p(a) \homeostrict p(f(a))$.  Other examples are $X
\homeo X$, $p(X) \homeostrict p(f(Y))$, $p(X,X) \homeo p(X,Y)$ and
$p(X,Y) \homeo p(X,X)$.

In order to adequately handle some built-ins,
 the embedding relation $\homeo$\index{$\homeo$}
\index{homeomorphic embedding!dynamic functors}
 of Definition~\ref{definition:homem} has to be adapted.
Indeed, some built-ins\index{built-ins} (like $=../2$ or $is/2$)
 can be used to dynamically construct 
new constants and functors. With an unbounded number of constants and
functors, $\homeo$ is not a wqo.
To remedy this \cite{LeuschelMartensDeSchreye:Toplas} partitions the
constants and functors into the {\em static\/} ones (those occurring
in the original program and the partial deduction query) and the {\em
  dynamic\/} ones (those created during program execution)\footnote{A
  similar division was used in {\sc mixtus} \cite{Sahlin93:ngc} to
  solve problems with subsumption checking.}.  As with the set of
variables, the set of dynamic constants and functors is unbounded.
Hence, not surprisingly a wqo is obtained by adding to
Definition~\ref{definition:homem} a rule similar to the rule for
variables:
\begin{quote}
 $f(s_{1},\ldots,s_{m}) \homeo g(t_{1},\ldots,t_{n})$
 if both $f$ and $g$ are dynamic
\end{quote}

\paragraph{{\bf Comparing wfos and wqos.}}

The homeomorphic embedding allows us to continue unfolding in situations
where no suitable wfo exists. For example, on its own 
 (i.e., not superimposed on a determinate unfolding strategy)
 it will allow the complete
unfolding of most terminating Datalog programs.
The homeomorphic embedding $\homeo$ allows also better unfolding in
the context of metaprogramming (see 
\cite{Leuschel:LOPSTR98,Vanhoof:PhD}).

Take for example the atoms
 $A = p([],[a])$
 and $B = p([a],[])$.
This is a situation where a homeomorphic embedding
allows more unfolding than any wfo:
 it allows us to unfold $A$
 when $B$ is in its covering ancestor sequence, but also the
other way around, i.e., it allows us to unfold $B$ when $A$ is
in its covering ancestor sequence. A wfo will at best assign a
different size to both atoms and the total order, fixed in advance
implies that only one of the two unfoldings can be performed.
The dynamic adjustment of wfos which we described in 
Example~\ref{example:refine-wfo}
 will allow both unfoldings.
However, if we make the above example
 slightly more complicated, e.g., by using the atoms
  $A=\mathit{solve}(p([],[a]))$
  $B=\mathit{solve}(p([a],[]))$ or even
  $A=\mathit{solve}_{1}(\ldots\mathit{solve}_{n}(p([],[a]))\ldots)$
  $B=\mathit{solve}_{1}(\ldots\mathit{solve}_{n}(p([a],[]))\ldots)$
  instead,
  then the scheme of Example~\ref{example:refine-wfo} will no longer work
  (while $\homeo$ still allows both unfoldings).
For such a wfo scheme to allow both unfoldings, we
 have to make the dynamic argument selection
  process more refined but then we run
 into the problem that infinitely many dynamic refinements
 might exists \cite{MartensDeSchreye:jlp95b,Martens:phd},
 and to our knowledge no satisfactory solutions exists as of yet.

However, the above example also illustrates why, when using a wqo, one has to
   compare with {\em every predecessor}.
Otherwise one will get infinite derivations
where in turn the atoms $p([a],[])$, $p([],[a])$ and again $p([a],[])$
are selected.
When using a wfo one has to compare only to the closest predecessor,
 because of the 
 transitivity of the order and the strict decrease enforced at each
 step.

Formally, one can prove that $\homeo$ is strictly more powerful
 than so-called 
 {\em simplification orderings\/} (such as lexicographic path
 ordering; see \cite{Dershowitz:htcs90}) and
 so-called {\em monotonic\/} wfos
 \cite{Leuschel:SAS98}:
 the admissible sequences with respect to $\homeo$ are
 a strict superset of the union
 of all admissible sequences with respect
 to simplification orderings and monotonic
wfos.
Almost all wfos presented in the online
  partial deduction literature
 so far fall into this category.
Also, compared to all these wfo-approaches,
the $\homeo$ approach is relatively easy to implement.
The combined power and simplicity
 explains its popularity in the recent years
 \cite{SorensenGluck:ILPS95,%
LeuschelMartensDeSchreye:Toplas,%
GlueckJorgensenMartensSorenson:control,%
JorgensenLeuschelMartens:CPPD,Alpuente:PEPM97,LafaveGallagher:LOPSTR97,%
AlbertEtAl:SAS98,VanHoofMartens:Parse,AlpuenteEtAl:TOPLAS98,%
AlbertEtAl:SAS98,CPD:megapaper}.

There are, however, natural wfos which
 are neither simplification orderings nor monotonic.
For such wfos, there can be
 sequences which are not admissible wrt $\homeo$ but which
  are admissible wrt the wfo.
Indeed, $\homeo$ takes the whole term structure into account
 while wfos in general can ignore part of the term structure.
For example, the sequence
 $\langle [ 1, 2 ] , [ [1,2] ]\rangle$ is admissible wrt 
 the ``listlength'' measure but not wrt $\homeo$,
  where ``listlength'' measures a term
  as 0 if it is not a list and by the number of elements in the list
  if it is a list \cite{MartensDeSchreye:jlp95b}.

In summary,
 the only circumstances when one might consider using wfos
  for online control instead of a
 wqo such as $\homeo$ are:
\begin{enumerate}
\item When the use of the wqo $\homeo$ is considered too inefficient
  (checking the extension of an admissible sequence for admissibility
  is much less expensive with a wfo than with a wqo).
\item When there is a need to consider only parts of the terms
  structures inside atoms. It is unclear how often 
 this is going to be important in practice.
  
\item When one wants to explicitly restrict the amount
  of unfolding, e.g., for pragmatic reasons.
\end{enumerate}

\subsection{Local control in {\sc ecce}}

Experience with {\sc ecce}, an {\bf online} partial deduction system
 \cite{Leuschel96:ecce-dppd},
has resulted in the following recommendations for unfolding a goal:
(the query is always unfolded, as needed for correctness): 
\begin{itemize}
\item If the goal fails (has a literal that does not unify with any
  clause head) then label the derivation as a failing one.

\item Else, try to find a determinate literal whose unfolding yields
  an SLDNF-derivation that is safe with respect to the wqo $\homeo$
  and unfold it. To decide whether a literal is determinate a lookahead of 1 is used.
  
\item Else, unfold the leftmost literal and stop with
  further unfolding of this branch (apart from identifying failing
  resolvents). This rule is not always giving the best unfolding. There
  are derivations where non-determinate unfolding is better omitted.
  Also it can be that the leftmost literal is a built-in or another
  literal that cannot be unfolded because its definition is not
  available. In such case, non-leftmost non-determinate unfolding can
  be considered if the amount of work duplication to be introduced is
  minimal (which is the case for cheap built-ins such as $\dif$) or
  will be minimised by a postprocessor or smart Prolog compiler.

\end{itemize}

These recommendations are not always sufficient. On benchmarks such as
the highly non-deterministic ``contains'' referred to in
Section~\ref{subsection:efficiencydeterminacy}, they are too restrictive.
Obtaining good specialisation requires to perform non-determinate
unfolding (and, as for determinate unfolding, it must be safe with respect
to the wqo $\homeo$). Interestingly, the default setting of {\sc ecce}
includes so-called ``conjunctive'' partial deduction (to be discussed
in Section~\ref{section:conjPD}) and determinate unfolding is
sufficient to handle ``contains'' and similar benchmarks.
The first version of {\sc ecce} described in
 \cite{LeuschelMartensDeSchreye:Toplas} did not include
 conjunctive partial deduction and thus non-determinate
 unfolding was employed.


\subsection{Termination within subsidiary SLDNF-trees}
\label{sec:subsidiary} 

In an SLDNF-derivation, there is not only the possibility of
non-termination for the main SLDNF-tree but also for all the
subsidiary SLDNF-trees. Under SLDNF, such subsidiary trees are only
created for ground atoms, hence their unfolding at specialisation-time
is not different from their execution at run-time.  However, as
control is different, some subsidiary trees can be created during
partial deduction which are never created at run-time.  Moreover, the
original program may be erroneous in the sense that the execution of
some of the subsidiary trees created at run-time does not terminate.
So, to ensure that the partial deduction of a program always
terminates, one has to control the execution of the subsidiary trees.

Non-termination can have two sources. On the one hand, an infinite
branch can be created. This is similar to the problem of creating an
infinite branch in the main tree, and the same local control
techniques can be used to prevent it. On the other hand, a ground
negative literal can be selected in a subsidiary tree, leading to the
creation of another subsidiary tree, and so on, eventually resulting into
an infinite set of subsidiary trees. This problem is similar to the
global termination problem  mentioned in
Section~\ref{section:basics:control} and can also be solved by the
same techniques (to be described in Section~\ref{section:global-control}).
Alternatively, one could conceptually attach
 the subsidiary trees to the main tree (i.e., when building a subsidiary
 tree for an atom $A$ we consider all childrens of $A$ also as childrens
 of $\neg A$ in the main tree) and then use the local control techniques
 which we discussed.

If the control interrupts the execution of the subsidiary tree before
it reports success or failure to the main node, then the negative atom
cannot be selected and the node becomes either an incomplete leaf or
another atom has to be selected.%
\footnote{In both cases the negative literal will feature in
 the residual program, and one should not throw the
  subsidiary trees away, as they can be used for code
  generation.}

\subsection{From pure logic programming to Prolog}
\label{impure}

\paragraph{Pure Prolog.}
As already mentioned, Theorem~\ref{theorem:pdcorrectness}
 guarantees neither that termination under, e.g.,  Prolog's
 left-to-right
 selection rule is preserved, nor that solutions are found in the same
order.
However, as shown in\cite{Proietti:PEPM91},
 there are further restrictions on the unfolding
 that can be imposed to remedy this 
 (and no further restrictions on the global control are necessary).
First, we have already seen
that determinate unfolding can only improve
 termination and will not change the order of solutions under Prolog.
Second, leftmost unfolding (determinate or not)
 changes  neither the termination nor the order of solution under
 Prolog execution.
Thus, if one prevents non-leftmost, non-determinate unfolding
 (as already discussed in Example~\ref{ex:inboth:duplication}
  this is also a good idea for efficiency)
  then partial deduction will
  always preserve termination (and could improve it)
  as well as the order of solutions for pure Prolog programs.

\paragraph{Full Prolog.}
So far we have only considered pure logic programs with declarative
built-ins (such as $\mathit{functor}$, $\mathit{arg}$,
$\mathit{call}$, cf.,
Example~\ref{ex:benefits-of-pd}).  
We were thus able to
 exploit the {\em independence of the selection rule\/}
\cite{Apt90:htcs,Lloyd:flp}, in the sense that the unfolding
 rule did not have to systematically select the leftmost literal in a goal.
We were thus able, e.g., to perform non-leftmost
 determinate unfolding steps (which can be the source of big speedups,
see \cite{LeuschelDeSchreye98:jlp}). In this section we briefly touch
upon the differences between partial deduction of pure logic programs
and partial evaluation of impure Prolog.

When we move towards full Prolog with extra logical built-ins, such as
$\mathit{var}$, the cut, or even $\mathit{assert}$, we can no longer make use of
the independence of the selection rule and our unfolding choices
become more limited as everything that modifies the procedural
semantics of the program may have an effect on the results computed by
it.

For the cut, the order of solutions is important, as the cut commits to
the first solution. Predicates such as $\mathit{nonvar}/1$ and
$\mathit{var}/1$ are what is called {\em binding-sensitive}. Success
or failure for e.g.\ $\mathit{var}(X), p(X)$ can be different than for
$p(X), \mathit{var}(X)$ and unfolding $p(X)$ in $\mathit{var}(X),
p(X)$ can result in so called {\em backpropagation} of bindings onto
the binding-sensitive call to $\mathit{var}/1$. Also the side effect
of a printing statement is binding-sensitive and backpropagation of a
failure may eliminate its execution altogether as in the
specialisation of $\mathit{print}(\mathit{hello}),$ $\mathit{fail}$
into $\mathit{fail}$.
Thus,
 any non-leftmost unfolding step, even when determinate, may cause a
change in the procedural semantics. 
Proposals to overcome this limitation can be found in, e.g.,
\cite{OKeefe:SLP85,BugliesiRusso:NACLP89,Prestwich:LOPSTR92,%
  Sahlin93:ngc,Sahlin:phd,Leuschel:LOPSTR94}.  In essence, one has to
avoid backpropagation of bindings onto binding-sensitive predicates.
For example, given a program $P$ containing a single fact $p(a)\larr$
for the predicate $p$, the goal $\larr \mathit{var}(X), q(X), p(X)$
(with $q$ not binding-sensitive) is specialised into $\larr
\mathit{var}(X),X\!=\!a,q(a)$. This avoids the backpropagation of $a$
into $\mathit{var}(X)$.

Similarly, one has to avoid backpropagation of failure onto predicates
with side-effects such as $\mathit{print}$.  E.g., for the same
program $P$ and a goal $\larr print(a),q(b)$, assuming all unfoldings
of $q(b)$ end in failure, one cannot specialise the goal into $\larr
fail$ but has to specialise it into $\larr print(a),fail$ instead.

A problem related to the cut is that unfolding an atom with a program
clause containing a cut modifies the scope of the cut: the SLDNF-tree
resulting from the execution of the specialised program is pruned
differently by the cut than the SLDNF-tree from the execution of the
original program. This problem is overcome by providing special
built-ins (mark-cut). They allow us to preserve the meaning of cut under
unfolding. The if-then-else, with its local cut, poses much less
problems and is preferable from a partial evaluation perspective
\cite{OKeefe:SLP85}.

Another problem relates to the specialisation of modules.  Some
systems (e.g., {\sc ecce} \cite{Leuschel96:ecce-dppd}) allow some predicates to
be annotated  as {\em open}. The specialiser assumes that
the definitions will be provided at runtime and does not unfold such
predicates. (For specialising Prolog, one should in addition declare
whether or not these predicates are binding-sensitive).
A solution for the G\"{o}del module system is presented in
 \cite{Gurr:PHD}, using the concept of a {\em script\/}
 where the module structure has basically been flattened.

 In summary, extending the control techniques to full Prolog is
 feasible.  In essence, one has to prevent the backpropagation of
 bindings, either by only performing leftmost unfolding or by some
 other means (e.g., the explicit introduction of equalities).
 However, as backpropagation can lead to early detection of failure
 and hence important speedups, it means that some interesting
 specialisations are no longer possible.  Figuring out, via some
 analysis, when a substitution can safely be backpropagated beyond a
 binding sensitive predicate call is a difficult challenge, and, to
 our knowledge, no satisfactory solution exists.
 

\section{Global Control}
\label{section:global-control}

\subsection{Independence and renaming/filtering}
\label{section:basics:renaming}

As we have seen in Section~\ref{chapter:basics},
 correctness of partial deduction
  requires that the atoms in $\pdset{A}$ are independent.
There are two ways to ensure the independence condition.  The first
one is to replace the atoms which are not independent by a more
general atom (first proposed in \cite{Benkerimi:naclp90}). For
example, replacing the dependent atoms $\mathit{member}(a,L)$ and
$\mathit{member}(X,[b])$ by $\mathit{member}(X,L)$ in a set
$\pdset{A}$ removes the dependency; moreover the new set is closed
with respect to all atoms in the old one. As discussed below, this
approach can also be used to ensure global termination. However, it
introduces precision loss as information about specific calls is
disregarded; hence it can worsen the degree of global specialisation.

A better way to address the independence problem uses a
 so-called  {\em
  renaming\/} transformation, which
 renames every atom of $\pdset{A}$ by
giving it a {\em distinct\/} predicate symbol; the set of atoms to
 be specialised thus becomes
independent {\em without\/} introducing any precision loss. For
instance, given the dependent atoms $\mathit{member}(a,L)$ and
$\mathit{member}(X,[b])$, renaming the second one into
$\mathit{member'}(X,[b])$ removes the independence. The renaming
transformation then also has to map the atoms inside the bodies of the
residual program clauses of $P'$ as well as atoms in queries for the
specialised program to the correct versions. For example it should
rename the query $\leftarrow
\mathit{member}(a,[a,c]),\mathit{member}(b,[b])$ into $\leftarrow
\mathit{member}(a,[a,c]),\mathit{member}'(b,[b])$.

Renaming can often be combined with so called argument filtering to
improve the efficiency of the specialised program.  The basic idea is
to filter out constants and functors and to keep only the variables as
arguments. In terms of the fold/unfold transformation framework
\cite{BurstallDarlington:ACM77,TamakiSato:iclp84,Pettorossi94:jlp} it
consists of defining new predicates and using it to fold occurrences
in $\pdset{A}$, $P'$, and $G$. Considering the same examples, defining
$\mathit{mem_a}(L) \leftarrow \mathit{member}(a,L)$ and
$\mathit{mem_b}(X) \leftarrow \mathit{member}(X,[b])$, the dependent
atoms $\mathit{member}(a,L)$ and $\mathit{member}(X,[b])$
are folded into the independent atoms  $\mathit{mem_a}([a,c])$
and $\mathit{mem_b}(b)$, while the query is folded into $\leftarrow
\mathit{mem_a}([a,c]),\mathit{mem_b}(b)$. 
Further details about filtering can be found in
 \cite{Gallagher91:ngc}, \cite{Benkerimi93:jlc},
 \cite{LeuschelSorensen:RAF} or \cite{Proietti93:jlp}.
The specialisations shown in  \cite{SafraShapiro:IFIP86} strongly
suggest that the authors already applied a form of argument filtering;
it has also been referred to as
 ``pushing down meta-arguments'' in \cite{SterlingBeer:jlp89}
 or ``PDMA'' in \cite{Owen:META88}.
In functional programming the term of ``arity raising''
 has also been used.
It has also been studied in an offline setting,
  where filtering is more complicated.
  
Renaming and filtering are used in a lot of practical approaches
 (e.g., 
\cite{Gallagher91:TR,Gallagher:PEPM93,Gallagher91:ngc,%
LeuschelDeSchreye:PEPM95,LeuschelDeSchreye98:jlp,%
LeuschelMartensDeSchreye:Toplas}) and adapted correctness results can
be found in \cite{Benkerimi93:jlc}.  To avoid the need for a renaming
transformation on queries to the specialised program, interface
predicates are provided that define the original predicates in terms
of the renamed ones.

   
\subsection{Syntax-based Global Control}
\label{section:basics:polyvariance}

Having solved the independence problem without introducing any
precision loss, we can now turn our attention to the problem of
ensuring {\em closedness\/} and {\em global termination\/} while
maximising the {\em degree of global specialisation}. In a so called
monovariant analysis, the problem is solved by keeping at most one
atom in $\pdset{A}$ for each predicate. When several atoms occur with
the same predicate symbol, they are replaced by a generalisation. This
ensures that each predicate has at most one specialised version,
ensuring correctness and ---as there are no infinite chains of
strictly more general expressions \cite{Huet:JACM80}---
termination. However, as already said, generalising atoms introduces
precision loss, hence it is worthwhile to consider {\em polyvariance},
the construction of several specialised versions of the same predicate.
Deciding exactly
 how many versions is referred to as the {\em control of
  polyvariance problem}. 

Let us examine how the closedness, global termination and the degree of
global specialisation interact:

\begin{zitemize}
\item {\em Closedness vs.\ Global Termination.}\\
 As we have seen in Procedure~\ref{algo:stdpd},
 closedness can be simply ensured by repeatedly adding the atoms
  which are not $\pdset{A}$-closed
 to $\pdset{A}$ and unfolding them.  Unfortunately this process (first
  presented in \cite{Benkerimi93:jlc}) is not guaranteed to terminate.
\item {\em Global Termination vs.\ Global Specialisation.}\\
  To ensure global termination one can use for the
  $\mathit{revise}$ function in Procedure \ref{algo:stdpd},
  a so-called
  {\em generalisation\/} operator, which generates a set of more general
  atoms. While replacing atoms by strictly more general ones
  introduces precision loss, it is sometimes essential to ensure
  termination. 

\end{zitemize}

The notion of generalisation can be formalised as follows:

\begin{definition} \label{def:abstract-std}
Let $\pdset{A}$ and $\pdset{A}'$ be sets of atoms.
Then $\pdset{A}'$ is a {\em generalisation\/}
 of $\pdset{A}$ iff every atom in $\pdset{A}$
  is an instance of an atom in $\pdset{A}'$.
 A {\em generalisation operator\/}
 is an operator which maps every finite set of atoms to a
 generalisation of it which is also finite.
\end{definition}

A generalisation operator is often referred to as
 an abstraction operator in the literature, but we think
the term generalisation is more appropriate.
 
With $\pdset{A}'$ a generalisation of $\pdset{A}$, 
 any set of clauses which is $\pdset{A}$-closed is also
 $\pdset{A}'$-closed.
Using a generalisation operator as $\mathit{revise}$ function in
Procedure \ref{algo:stdpd} does not guarantee global termination.
But, if the procedure terminates then closedness is ensured, i.e., $P'
\cup \{S\}$ is $\pdset{A}$-closed (modulo renaming).  With this
observation we can reformulate the {\em control of polyvariance
  problem\/} as one of finding a {\em generalisation operator which
  maximises the global degree of specialisation
  while ensuring termination}.
In the rest of this section we will survey methods that only consider
the syntactic structure of the atoms to be specialised.


\subsubsection{Most specific generalisation}

\begin{definition} The {\em most specific
    generalisation\/} or {\em least general generalisation} of a
  finite set of expressions $E$, denoted by $\mathit{msg}(E)$, is the
  most specific expression $M$ such that all expressions in $E$ are
  instances of $M$.
\end{definition}

\begin{figure}[htb]
\begin{center}\begin{small}
\begin{tabular}{@{}lll}
$A$ & $B$ & $\mathit{msg}(\{A,B\})$\\ \hline
$a$ & $b$ & $X$\\
$p(a,b)$ & $p(a,c)$ & $p(a,X)$\\
$p(a,a)$ & $p(c,c)$ & $p(X,X)$\\
$p(0,s(0))$ & $p(0,s(s(0)))$ & $p(0,s(X))$\\
$q(0,f(0),0)$~~ & $q(a,f(a),f(a))$~~ & $q(X,f(X),Y)$~~\\
$r(a)$ & $r(s(a))$ & $r(X)$\\
\end{tabular}
\end{small}\end{center}
\caption{Examples of {\em msg}\label{fig:msg}}
\end{figure}

Some examples can be found in Figure~\ref{fig:msg}.  The msg can be
effectively computed \cite{LassezMaherMarriott:fddlp88}. The algorithm
is also known as
{\em
anti-unification}.
and dates back to \cite{Plotkin:MI69} and 
 \cite{Reynolds:MI69}.
As already mentioned, giving an expression $A$, 
there are no infinite chains of strictly more general expressions
\cite{Huet:JACM80}.

This makes the msg well suited for use in a generalisation operator.
One of the first generalisation operators was proposed in
\cite{Benkerimi:naclp90}. It applied the msg on atoms which have a
common instance.  As first pointed out in
\cite{MartensDeSchreyeHorvath:TCS94}, this does not ensure
termination, as can be seen when specialising Example~\ref{ex:revacc}
for the initial goal $\larr \mathit{rev}(X,[],R)$ (no matter which
terminating unfolding rule is used, all atoms in $\pdset{A}'_i$ are
independent, hence
$\mathit{generalise}(\pdset{A}'_i) = \pdset{A}'_i$ and the set
is growing forever).

A simple generalisation operator which ensures termination is obtained
by imposing a finite maximum number of atoms in $\pdset{A}_{i}$ for
each predicate and using the msg to stick to that maximum (e.g.\ 
\cite{MartensDeSchreyeHorvath:TCS94}. However, the msg introduces
precision loss and is applied at an arbitrary point.  As illustrated
in \cite{MartensDeSchreyeHorvath:TCS94}, there will be cases where the
msg is applied too early and precision loss is introduced that should
have been avoided; in other cases, the msg is applied too late,
resulting in too many uninteresting variants and code explosion.

\subsubsection{Global Trees with wfos and wqos}
\label{subsection:global-trees}

We therefore need a more principled approach to global termination,
much as we needed a more principled approach to local termination in
Section~\ref{section:basics:unfolding}.  Probably the first such
solution, not depending on any ad-hoc bound, is
\cite{MartensGallagher:ICLP95}.  The idea is to use the wfo approach
also to ensure global termination.  To do this,
\cite{MartensGallagher:ICLP95} proposed to structure the current atoms
$\pdset{A}_i$ (see Procedure~\ref{algo:stdpd}) to be partially deduced
as a so-called {\em global tree}: i.e., a tree whose nodes are labeled
by atoms and where $A$ is a child of $B$ if specialisation of $B$
leads to the specialisation of $A$, in the sense that
 $A\in \mathit{leaves}(\mathit{unfold}(P,B))$.
This gives us a structure very
similar to the SLDNF-trees encountered by the local control, and thus
enables to apply wfo in much the same manner.  In
\cite{LeuschelMartensDeSchreye:Toplas}, this was extended to also
accommodate wqos (and characteristic trees; which we discuss later).

Figure~\ref{figure:procedure2}
 contains a generic procedure based upon
\cite{MartensGallagher:ICLP95,LeuschelMartensDeSchreye:Toplas}.

\begin{figure}
\begin{procedurerm} 
\label{alg:basictree}
\noindent
{\bf Input:} a program $P$ and a set $S$ of atoms of interest;

\noindent
{\bf Output:} A specialised program $P'$ and a set of atoms
$\pdset{A}$;

\noindent
{\bf let} ${\gamma}=$ a ``global'' tree consisting of a
   marked unlabeled root node $R$;\\
{\bf for } each $A \in S$ {\bf do}
  \begin{zitemize}
    \item[] create in $\gamma$ a new unmarked node $C$ as a child of $R$;
   \item[] {\bf let} $\mathit{label}(C)$ := $A$
  \end{zitemize} 

\noindent
{\bf repeat}
  \begin{zitemize}
  \item[] {\bf pick} an unmarked leaf node $N$ in ${\gamma}$; 
  \item[] {\bf if} $\mathit{covered}(N,\gamma)$ {\bf then} mark $N$ as covered
  \item[] {\bf else}
  \item[~~~~] {\bf let} $W = \mathit{whistle}(N,\gamma)$;
  \item[~~~~] {\bf if} $W \neq \mathit{fail}$ {\bf then}
    {\bf let} $\mathit{label}(N)$ :=  $\mathit{generalise}(N,W,\gamma)$%
    \footnote{The ECCE system allows the user to select an
      alternative approach that removes all descendants of
      $W$ and changes the label of $W$.}
   \item[~~~~] {\bf else}
   \item[~~~~~~~~] mark $N$ as processed
   \item[~~~~~~~~] {\bf for all} atoms 
     $A \in \mathit{leaves}(\mathit{unfold}(P,\mathit{label}(N)))$ {\bf do}
   \item[~~~~~~~~~~~~]  create in $\gamma$ a new unmarked node $C$ as a child of $L$;
   \item[~~~~~~~~~~~~] {\bf let} $\mathit{label}(C)$ := $A$
  \end{zitemize}
{\bf until} all nodes are marked;

\noindent
{\bf let} $\pdset{A}:= \{\mathit{label}(N) \mid N\in \gamma \mbox{ and
  } N \mbox{ is not marked as covered}\}$;\\
{\bf let} $P' := \bigcup_{A\in
  \pdset{A}} \mathit{resultants}(\mathit{unfold}(P,A))$ 
\end{procedurerm}
        \caption{Generic tree-based partial deduction procedure}
        \label{figure:procedure2}
\end{figure}

The procedure is parameterised by the unfold function
$\mathit{unfold(P,A)}$, the predicate $\mathit{covered}(N,\gamma)$,
the whistle function $\mathit{whistle}(N,\gamma)$ and the generalisation
function $\mathit{generalise}(N,W,\gamma)$. The unfold function takes
care of the local control and returns a finite SLDNF-tree. The
predicate $\mathit{covered}(N,\gamma)$ decides whether there is
already a partial deduction suitable for the atom $\mathit{label}(N)$.
Termination and
 correctness require that it must return $\mathit{true}$ when
there is another marked node in the same branch labelled with a variant of
$\mathit{label}(N)$ and that, whenever it returns $\mathit{true}$, the
global tree $\gamma$ has a marked node $M$ such that
$\mathit{label}(M)\theta = \mathit{label}(N)$ for some substitution
$\theta$. The whistle function $\mathit{whistle}(N,\gamma)$ prevents
the growth of infinite branches in the global tree by using wfos or wqos; 
it raises an alarm
by returning an ancestor node $W$ of $N$ in case $N$ is not an
 admissible descendant of $W$ 
(hence $\mathit{label}(W)$ has the same predicate symbol as 
$\mathit{label}(N)$)
and $\mathit{fail}$ otherwise. If $N$ is
not admissible, it has to be generalised. 
The generalisation function  $\mathit{generalise}(N,W,\gamma)$ computes a
generalisation of $\mathit{label}(N)$. To ensure termination, it must be 
a strict generalisation. 
Besides $N$ it takes as parameters $W$ and $\gamma$. The
latter allows the function to
return a generalisation that is admissible with respect to the whole
branch ending in $N$. 
As the generalisation can now be covered by another marked node of the
global tree, $N$ should not yet be marked. If $N$ is admissible, its
label is unfolded and the leaves of the obtained SLDNF-tree are added as
unmarked children of $N$ while $N$ is marked. Once all nodes are marked,
the set $\pdset{A}$ and the specialised program are extracted.

Observe that in the above procedure the generalisation operator of 
 Definition~\ref{def:abstract-std}
 is split up into three components
 $\mathit{covered}(N,\gamma)$, 
 $\mathit{generalise}(N,W,\gamma)$, and $\mathit{whistle}(N,\gamma)$.
An instantiation of these three components that ensures correctness
and terminations and uses the wqo $\homeo$ for $\mathit{whistle}(N,\gamma)$
is as follows 
(this is one of the possible settings in {\sc ecce}):
\begin{zitemize}
 \item $\mathit{whistle}(N,\gamma)$ =
   $W$  iff $W$ is the closest ancestor
of $L$ such that $\mathit{label}(W) \homeo \mathit{label}(L)$
 and $\mathit{label}(L)$ is not
 strictly more general than $\mathit{label}(W)$,%
\footnote{This latter test is required to avoid some technical
 difficulties with the way $\homeo$ treats variables;
 see \cite{LeuschelMartensDeSchreye:Toplas,Leuschel:LOPSTR98}.};\\
$\mathit{whistle}(L,\gamma)$ = $fail$ if there is no such ancestor.
\item $\mathit{generalise}(N,W,\gamma)$ =
 $\mathit{msg}(\mathit{label}(N),\mathit{label}(W))$
 \item $\mathit{covered}(N,\gamma)$ = $\mathit{true}$
 if there is a node $M$ in $\gamma$ such that $\mathit{label}(M)$ is a variant of
 $\mathit{label}(N)$;\\
 $\mathit{covered}(N,\gamma)$ = $\mathit{false}$ otherwise.
\end{zitemize}

\paragraph{Discussion}
There are a few works within partial deduction of logic programs,
 in which the local and global control interact much more tightly,
 in the sense that the local control also takes information from
 the global control into account
\cite{Sahlin93:ngc,GlueckJorgensenMartensSorenson:control,CPD:megapaper,%
VanHoofMartens:Parse}.
Also observe that, in other
 programming paradigms such as supercompilation of functional
 languages
 \cite{Turchin:toplas86,GlueckSorensen:PE96,SorensenGlueckJones:jfp,%
Sorensen:PE98}, historically there has not been a
 clear distinction between
 local and global control. In these settings,
 e.g., \cite{SorensenGluck:ILPS95,SorensenGlueckJones:jfp,Sorensen:MPC98}
 there is only one big ``global'' tree which is then
 cut up into local trees during the code generation. This approach is
 also taken in the ``compiling control'' transformation of logic
 programs in 
 \cite{BruynoogheDeSchreyeKrekels:jlp91}.
In the future, it might be interesting to compare these
 two approaches systematically from a pragmatic point of view.

\subsection{Computation-based Global Control}
\label{chapter:chtree}

\subsubsection{Characteristic trees}
\label{sect:chtree:chpaths-trees}

While the global trees of Section \ref{subsection:global-trees} show
the relationship between roots and leaves of constructed SLDNF-trees,
the generalisation function which generalises the atoms is purely
syntactical. It only takes into account the atoms as they appear in
the global tree.  However, the
same two atoms can behave in a very similar way in the context of one
program $P_1$, but in a very dissimilar fashion in the context of
another program $P_2$.  The syntactic structure of the two atoms being
unaffected by the particular context, the generalisation function 
$\mathit{generalise}(N,W,\gamma)$ will
thus perform exactly the same generalisation%
\footnote{Note, however, that $\mathit{whistle}(N,\gamma)$
  can behave differently as $\gamma$ will have a different structure.}
 within $P_1$ and $P_2$, even
though very different action might be called for.  A much more
appealing approach, might therefore be to examine the SLDNF-trees
generated for these atoms.  These trees capture (to some depth) how
the atoms behave computationally in the context of the respective
programs.  They also depict the specialisation that has been performed
on these atoms.  A generalisation operator which takes these trees into
account will notice their similarity in the context of $P_1$ and their
dissimilarity in $P_2$, and can therefore take appropriate actions in
the form of different generalisations.

This observation lead to the definition of {\em characteristic
  trees\/}, initially presented in
\cite{Gallagher91:ngc,Gallagher91:TR} and later exploited in
\cite{LeuschelDeSchreye97:ngc,LeuschelMartensDeSchreye:Toplas}.  In
essence, characteristic trees abstract SLDNF-trees by only
remembering, for the non-failing branches:%
\begin{enumerate}
\item The position of the selected literals.
\item An identification of the clauses $C_1$, $C_2$, \ldots used in
  the SLDNF-derivation of the branch.
\end{enumerate}

\begin{figure}[htb]
\begin{center}
\setlength{\unitlength}{0.007in}%
\begin{small}
\begin{picture}(488,155)
\thicklines
              \put(124,62){\vector(0,-1){32}}
              \put(82,125){\vector(1,-1){39}}
              \put(71,125){\vector(-1,-1){39}}
              \put(335,125){\vector(-1,-1){39}}

\thinlines    \put(389,62){\line(0,-1){32}}
              \put(377,13){{\em fail}}
              \put(280,70){$\Box$}
              \put(338,70){$\larr \underline{\mathit{arc}(b,Y)},$
                            $\mathit{path}([Y|T])$}
              \put(346,125){\vector(1,-1){39}}
              \put(301,136){$\larr \underline{\mathit{path}([b|T])}$}
              \put(89,17){$\larr \mathit{path}([b|T])$}
              \put(60,70){$\larr \underline{\mathit{arc}(X,Y)},$
                             $\mathit{path}([Y|T])$}
              \put(16,70){$\Box$}
              \put(44,136){$\larr \underline{\mathit{path}(L)}$}
\end{picture}\end{small}
\caption{SLD-trees for Example~\protect\ref{ex:graph}
 \label{figure:ex:graph}}
\end{center}
\end{figure}
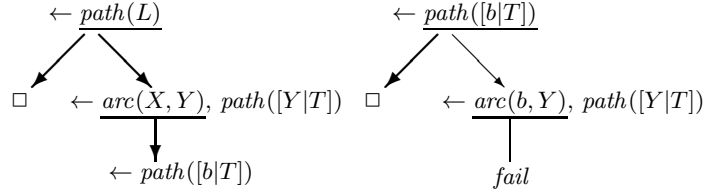

\noindent
We use $\ce{pos}{cl}$ to denote a derivation step that selects a
literal at position $pos$ and uses the clause identified by $cl$ to
compute a resolvent. A derivation or branch is represented as a
sequence of derivation steps and a characteristic tree as a set of
branches.
The information in a characteristic tree is sufficient to rebuild the
whole SLDNF-tree, hence it represents, directly or indirectly, all
successful, failing and incomplete derivations.
Two atoms with the same characteristic tree have so much in common
(same number and ``shape'' of residual clauses) that one would expect
that the same residual clauses can be used for both. We will discuss
below whether and how that can be achieved. First we look at an example
 which shows that characteristic trees can also be
 useful for the whistle function $\mathit{whistle}(N,\gamma)$:

\begin{example} \label{ex:graph}
Let $P$ be the following definite program:
\begin{enumerate}
\item[(1)]      $\mathit{path}([N]) \leftarrow$
\item[(2)]      $\mathit{path}([X,Y|T]) \leftarrow
\mathit{arc}(X,Y),\mathit{path}([Y|T])$
\item[(3)]      $\mathit{arc}(a,b) \leftarrow$
\end{enumerate}

Unfolding $\larr \mathit{path}(L)$ (e.g., using an unfolding rule based on
 $\homeo$; see Figure~\ref{figure:ex:graph} for the SLD-trees
 constructed)
 will result in lifting $\mathit{path}([b|T])$ to the global level.
Notice that we have a growth of syntactic structure
 ($\mathit{path}(L) \homeo \mathit{path}([b|T])$).
However, one can see that further unfolding $\mathit{path}([b|T])$
 results in an SLD-tree whose characteristic tree
 $\tau_B = \{\cpbeg\ce{1}{1}\cpend\}$ is strictly smaller than the
 one for $\mathit{path}(L)$
 (which is $\tau_A = $
 $\{\cpbeg\ce{1}{1}\cpend,\cpbeg\ce{1}{2}\cpdel\ce{1}{3}\cpend\}$).
\end{example}

As the example  illustrates
the {\em growth\/} of syntactic structure
can be accompanied by a {\em shrinking\/} of the associated
 SLDNF-trees.
In such situations
 there is, despite the growth of syntactic structure,
 actually no danger of non-termination.
A whistle function solely focussing on the syntactic structure
 would unnecessarily force generalisation,
 possibly resulting in 
a loss of precision.
Other examples can
 be found in \cite{LeuschelMartensDeSchreye:Toplas}.

Incorporating characteristic trees into the global control
 has 
 proven to be an elegant solution to avoid over-generalisation
 in several circumstances
 (when specialising meta-interpreters
  \cite{Leuschel:phd,VanHoofMartens:Parse} or
 when specialising pattern matchers to
 obtain the ``Knuth-Morris-Pratt'' effect \cite{Sorensen:PE98}).

A straightforward use of
 characteristic trees is as follows: classify
atoms at the global control level by their associated characteristic
tree and apply generalisation ($\mathit{msg}$) only on those atoms
which have the same characteristic tree.  This is basically the
approach pursued in \cite{Gallagher91:ngc,Gallagher91:TR}.
Unfortunately, the approach 
has some problems.
First, generalisation induces precision loss, even to the
extent that the generalised atom has a different characteristic tree.
Second, in case the number of distinct characteristic trees
 is not bounded, this approach will not terminate.
We illustrate these two problems, and how to remedy them, in the
 next two subsections.
 
\subsubsection{Preserving characteristic trees upon generalisation}
\label{section:preserve-chtree}

\begin{example} \label{ex:nonpreservation}
Let $P$ be the program:

\begin{small}\begin{pitemize}
\item (1)~~~~~$p(X)\leftarrow q(X)$
\item (2)~~~~~$p(c)\leftarrow$
\end{pitemize}\end{small}

Let $\pdset{A} = \{p(a),p(b)\}$.  Assume that $q(X)$ is not unfolded.
The atoms $p(a)$ and $p(b)$ have the same characteristic tree $\tau$ =
$\{\cpbeg\ce{1}{1}\cpend\}$.  Their msg, the atom $p(X)$ has a
different characteristic tree, namely
$\tau'=$ $\{\cpbeg\ce{1}{1}\cpend,\cpbeg\ce{1}{2}\cpend\}$ $\neq \tau$
and the specialisation for the atoms $p(a)$ and $p(b)$, due to the
inapplicability of clause (2), is lost in the partial deduction of
$p(X)$.  More importantly, there exists {\em no\/} atom, more general
than $p(a)$ and $p(b)$, which has $\tau$ as its characteristic tree.
\end{example}

The problem is that derivations that were absent in the original
characteristic trees appear in the characteristic tree of the
generalised atom. With negative literals, another source of difference
is that a negative literal, ground (and selected) at some point in the
original derivation is not necessarily ground, hence cannot be
selected, in the SLDNF-tree of the generalised atom.
More realistic examples can be found in
 \cite{LeuschelMartensDeSchreye:Toplas,LeuschelDeSchreye97:ngc}.

Two different solutions to this problem are:

\begin{enumerate}
\item {\em Ecological Partial Deduction.}
  \cite{Leuschel:LOPSTR95,LeuschelMartensDeSchreye:Toplas}
  
  The basic idea is to use the characteristic tree as a recipe to
  build part of the SLDNF-tree (and to ignore the part not constructed
  by following the recipe). In Example~\ref{ex:nonpreservation}, it
  means that the atom $p(X)$ is selected and clause (1) is used to
  construct a resolvent but that clause (2) is discarded as the branch
  using clause (2) is missing from the characteristic trees of $p(a)$
  and $p(b)$. Extracting the residual clauses from the part of the
  SLDNF-tree that has been built yields the clause $p(X)\leftarrow
  q(X)$.

  The pruning possible for $p(a)$ and $p(b)$ 
  is now preserved.  However, the residual code is not correct for all
  instances of $p(X)$; it is only correct for those instances
  for which $\tau$ is a possible characteristic tree. Hence, in
  Algorithm \ref{alg:basictree}, the function
  $\mathit{covered}(N,\gamma)$ should return $\mathit{true}$ only if
  there is a node $M$ such that $\mathit{label}(N)$ is an instance of
  $\mathit{label}(M)$ and if both have the same characteristic tree.
  In the example, the residual clause is correct for $p(a)$, $p(b)$,
  $p(d)$, but {\em neither\/} for $p(c)$ nor for $p(X)$.
Note that this approach also works with negative selected
 literals, and the
 above $\mathit{covered}(N,\gamma)$ test ensures that these negative
 literals do not become non-ground for the instances.
 
\item {\em Constrained Partial Deduction.}
  \cite{LeuschelDeSchreye97:ngc,LafaveGallagher:LOPSTR97}

  Whereas in standard partial deduction the members of $\pdset{A}$
  hence the roots of the SLDNF-trees are atoms, in constrained partial
  deduction, they are constrained atoms of the form $C\cbox A$, where
  $A$ is an atom and $C$ a constraint over some domain ${\cal D}$ (see
  \cite{Jaffar94:jlp} for details on constraint logic programming).
 \cite{LeuschelDeSchreye97:ngc} use
  inequality constraints over the Herbrand universe. Considering again
  the generalisation of the characteristic trees for the atoms $p(a)$
  and $p(b)$ of Example~\ref{ex:nonpreservation}, they derive as
  generalisation the constrained atom $X\neq c\cbox p(X)$. This atom
  has the same characteristic tree as the original atoms.
 This also requires the $\mathit{covered}(N,\gamma)$ to be adapted,
namely to check constraint entailment. However, constraints only
 appear during the partial deduction phase and the final specialised
  program is a pure logic program without constraints.
  Finally, this approach does 
 not allow us to select negative literals, but is more powerful
 than the ecological partial deduction approach for definite
 programs, as the derived constraints are not just used locally
 to obtain the desired characteristic tree but they can be propagated 
 globally to 
 other atoms in $\pdset{A}$ as well.

\end{enumerate}

Recently, {\em trace terms\/} have also been used in place of
characteristic trees \cite{GallagherLafave:PE96}.  Trace terms
abstract away from the particular selection rule, making them more
appealing in the context of pure logic programs.  They also have the
effect of 
providing a recipe during specialisation
thus achieving the effect of
ecological partial deduction, and they are easier to generate when
using the cogen approach \cite{MartinLeuschel:PSI99,Martin:phd}.

\subsubsection{Ensuring termination without depth-bounds}
\label{section:remove-depthbounds}

It turns out that for a fairly {\em large class of realistic
  programs\/} (and unfolding rules), the characteristic tree based
approaches described above only terminate when imposing a depth bound
on characteristic trees.
As the following simple example shows, this can
 lead to undesired
results when the depth bound is actually required.

\begin{figure}[tp]
\begin{center}
\setlength{\unitlength}{0.008in}%
\begin{picture}(488,382)(10,445)
\thicklines
\put( 60,810){\vector( 1,-1){ 30}}
\put( 50,810){\vector(-1,-1){ 30}}
\put( 95,755){\vector( 0,-1){ 30}}
\put( 30,795){\makebox(0,0)[rb]{\raisebox{0pt}[0pt][0pt]{\scriptsize (1)}}}
\put( 15,765){\makebox(0,0)[b]{\raisebox{0pt}[0pt][0pt]{\footnotesize
$\Box$}}}
\put(100,795){\makebox(0,0)[rb]{\raisebox{0pt}[0pt][0pt]{\scriptsize (2)}}}
\put( 55,819){\makebox(0,0)[b]{\raisebox{0pt}[0pt][0pt]{\footnotesize
 $\leftarrow \underline{\mathit{rev}(L,[],R)}$}}}
\put(105,765){\makebox(0,0)[b]{\raisebox{0pt}[0pt][0pt]{\footnotesize
 $\leftarrow \underline{\mathit{ls}([])},\mathit{rev}(T,[H],R)$}}}
\put( 95,705){\makebox(0,0)[b]{\raisebox{0pt}[0pt][0pt]{\footnotesize
 $\leftarrow \mathit{rev}(T,[H],R)$}}}
\put(120,740){\makebox(0,0)[rb]{\raisebox{0pt}[0pt][0pt]{\scriptsize (3)}}}
\put( 60,660){\vector( 1,-1){ 30}}
\put( 50,660){\vector(-1,-1){ 30}}
\put( 95,605){\vector( 0,-1){ 30}}
\put( 95,545){\vector( 0,-1){ 30}}
\put( 30,645){\makebox(0,0)[rb]{\raisebox{0pt}[0pt][0pt]{\scriptsize (1)}}}
\put( 15,615){\makebox(0,0)[b]{\raisebox{0pt}[0pt][0pt]{\footnotesize
$\Box$}}}
\put(100,645){\makebox(0,0)[rb]{\raisebox{0pt}[0pt][0pt]{\scriptsize (2)}}}
\put( 55,669){\makebox(0,0)[b]{\raisebox{0pt}[0pt][0pt]{\footnotesize
 $\leftarrow \underline{\mathit{rev}(T,[H],R)}$}}}
\put(125,615){\makebox(0,0)[b]{\raisebox{0pt}[0pt][0pt]{\footnotesize
 $\leftarrow \underline{\mathit{ls}([H])},\mathit{rev}(T',[H',H],R)$}}}
\put( 95,500){\makebox(0,0)[b]{\raisebox{0pt}[0pt][0pt]{\footnotesize
 $\leftarrow \mathit{rev}(T',[H',H],R)$}}}
\put(120,530){\makebox(0,0)[rb]{\raisebox{0pt}[0pt][0pt]{\scriptsize (3)}}}
\put(125,555){\makebox(0,0)[b]{\raisebox{0pt}[0pt][0pt]{\footnotesize
 $\leftarrow \underline{\mathit{ls}([])},\mathit{rev}(T',[H',H],R)$}}}
\put(120,590){\makebox(0,0)[rb]{\raisebox{0pt}[0pt][0pt]{\scriptsize (4)}}}
\put(355,780){\vector( 1,-1){ 30}}
\put(345,780){\vector(-1,-1){ 30}}
\put(390,725){\vector( 0,-1){ 30}}
\put(390,605){\vector( 0,-1){ 30}}
\put(390,665){\vector( 0,-1){ 30}}
\put(354,830){\makebox(0,0)[b]{\raisebox{0pt}[0pt][0pt]{\footnotesize In
general:}}}
\put(325,765){\makebox(0,0)[rb]{\raisebox{0pt}[0pt][0pt]{\scriptsize (1)}}}
\put(310,735){\makebox(0,0)[b]{\raisebox{0pt}[0pt][0pt]{\footnotesize
$\Box$}}}
\put(395,765){\makebox(0,0)[rb]{\raisebox{0pt}[0pt][0pt]{\scriptsize (2)}}}
\put(415,710){\makebox(0,0)[rb]{\raisebox{0pt}[0pt][0pt]{\scriptsize (4)}}}
\put(350,789){\makebox(0,0)[b]{\raisebox{0pt}[0pt][0pt]{\footnotesize
 $\leftarrow \underline{\mathit{rev}(T,\overbrace{[...]}^{n},R)}$}}}
\put(420,735){\makebox(0,0)[b]{\raisebox{0pt}[0pt][0pt]{\footnotesize
 $\leftarrow \underline{\mathit{ls}([...])},\mathit{rev}(T',[H',...],R)$}}}
\put(390,550){\makebox(0,0)[b]{\raisebox{0pt}[0pt][0pt]{\footnotesize
 $\leftarrow \mathit{rev}(T',[H',...],R)$}}}
\put(415,585){\makebox(0,0)[rb]{\raisebox{0pt}[0pt][0pt]{\scriptsize (3)}}}
\put(420,620){\makebox(0,0)[b]{\raisebox{0pt}[0pt][0pt]{\footnotesize
 $\leftarrow \underline{\mathit{ls}([])},\mathit{rev}(T',[H',...],R)$}}}
\put(415,650){\makebox(0,0)[rb]{\raisebox{0pt}[0pt][0pt]{\scriptsize (4)}}}
\put(390,680){\makebox(0,0)[b]{\raisebox{0pt}[0pt][0pt]{\footnotesize ...}}}
\put(400,677){$\left.\begin{array}{c}~\\~\\~\\~\end{array}\right\}$
                              \footnotesize $n$}
\end{picture}
\caption{SLD-trees for Example~\protect\ref{ex:revtype}.
\label{figure:revtype}}
\end{center}
\end{figure}
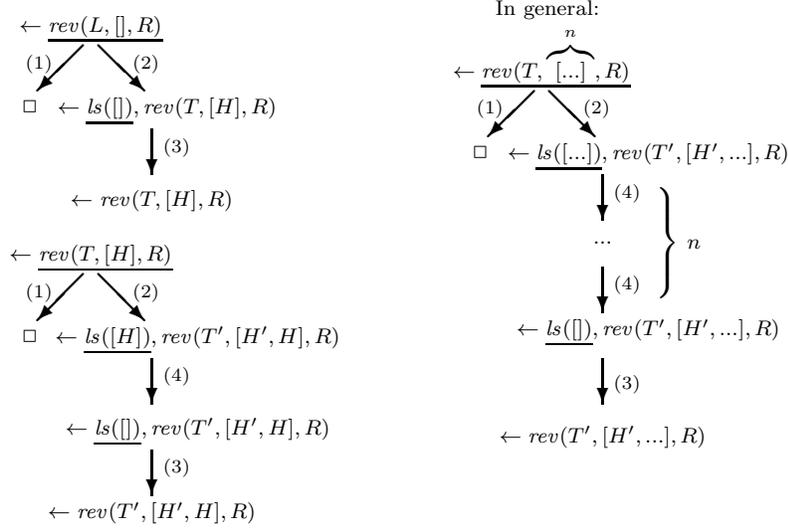

\begin{example} \label{ex:revtype}
  A list type check
  on the second argument (the ``accumulator'') is added to the reverse
  program from Example~\ref{ex:revacc}
\begin{enumerate}
\item[(1)]      $\mathit{rev}([],\mathit{Acc},\mathit{Acc}) \leftarrow$
\item[(2)]      $\mathit{rev}([H|T],\mathit{Acc},\mathit{Res}) \leftarrow$
 $\mathit{ls}(\mathit{Acc}),\mathit{rev}(T,[H|\mathit{Acc}],\mathit{Res})$
\item[(3)]      $\mathit{ls}([]) \leftarrow$
\item[(4)]      $\mathit{ls}([H|T]) \leftarrow \mathit{ls}(T)$
\end{enumerate}
As can be noticed in Figure~\ref{figure:revtype}, by using, e.g.,
determinate, $\homeo$-based, or well-founded unfolding we obtain
 an infinite number of different
 atoms, all with a different characteristic tree.
Imposing a depth bound of say 100, we obtain termination; however,
 100 different characteristic trees (and instantiations of the
 accumulator) arise, and 100 different
 versions of $\mathit{rev}$ are generated: one for each characteristic tree.
The resulting
 specialised program is certainly far from optimal and clearly
exhibits the ad hoc nature of the depth bound.
\end{example}

Situations like the above typically arise when some argument is
growing with the level of recursion and when this argument has an
influence on the characteristic tree of the SLDNF-tree built by the
$\mathit{unfold}$ function.
With simple programs such as Example~\ref{ex:revacc}, the growing
argument has no effect on the characteristic tree and it was believed
for some time that the problem would not arise in ``natural'' logic
programs.
However, among larger and more sophisticated programs, cases like the
above become more and more frequent, even in the absence of
type-checking. 

A solution to this problem is developed in
 \cite{LeuschelMartensDeSchreye:Toplas}, whose basic ingredients are as
follows:
\begin{enumerate}
\item Register descendancy relationships among atoms and their
  associated characteristic trees at the global level, by putting them
  into a {\em global tree} (as in Section~\ref{subsection:global-trees}).
\item Watch over the evolution of the characteristic trees associated
  with atoms along the branches of the global tree in order to detect
  inadmissible branches (as in Section~\ref{subsection:global-trees}).
  As suggested by Figure~\ref{figure:revtype}, a measure is needed
  that can spot when a characteristic tree (piecemeal) ``contains''
  characteristic trees appearing earlier in the same branch of the
  global tree.  
  An extension of the homeomorphic embedding relation
    can be used for this
     \cite{LeuschelMartensDeSchreye:Toplas}.
  If such a situation arises---as it indeed does in
  Example~\ref{ex:revtype}---one stops expanding the
  global tree, generalises the offending atoms, and produces a
  specialised procedure for the generalisation instead. 
  Note that in this case, it is actually impossible to preserve the
   characterstic trees upon generalisation, as the offending atoms
    will have different characteristic trees.
\end{enumerate}

The techniques formally elaborated in
\cite{LeuschelMartensDeSchreye:Toplas} have led to the implementation
of the {\sc ecce} system \cite{Leuschel96:ecce-dppd}.  The {\sc ecce}
system also handles (declarative) Prolog built-ins; these are also
registered within the characteristic trees (see \cite{Leuschel:phd}).


\section{Conjunctive Partial Deduction and Unfold/Fold}
\label{section:conjPD}

\subsection{Principles}

Partial deduction, as defined above
 (i.e., based upon the Lloyd-Shepherdson framework
 \cite{Lloyd:jlp91}),
 specialises a {\em set of atoms}.
Even though
 conjunctions of literals may appear within the SLDNF-trees constructed for these atoms,
 only atoms are allowed to appear at the global level.
In other words, when we stop unfolding, every conjunction at the leaf
 is automatically
 split into its atomic constituents which are then specialised
 (and possibly further abstracted) separately at the global level.
This restriction often
 considerably restricts the potential power of partial deduction,
 e.g., preventing the elimination
 of unnecessary variables \cite{Proietti:PLILP91}
 (also called 
 deforestation and tupling).

To overcome this limitation,
\cite{LeuschelDeSchreyeDeWaal:JICSLP96,%
GlueckJorgensenMartensSorenson:control,Leuschel:phd} present a
 relatively small extension of partial
deduction, called {\em conjunctive partial deduction}.  This technique
extends the standard partial deduction approach by 
considering sets $S = \{ C_1 , \ldots, C_n \}$ 
where the elements $C_i$ are now {\em conjunctions\/} of atoms
 (to some extent negative literals can also be used within conjunctions)
 instead of just single atoms.

Now, as the SLDNF-trees constructed for each $C_{i}$
are no longer restricted to having
{\em atomic} top-level goals,
 resultants (cf.\ Definition~\ref{def:basics:resultant-derivation})
 are not necessarily Horn clauses anymore: their left-hand side may
 contain a conjunction of literals.
To transform such resultants back into standard clauses,
 conjunctive partial deduction requires a {\em renaming\/} transformation,
 from conjunctions to atoms, in a post-processing step.
As with argument filtering, it can be formalised in the fold/unfold
 transformation framework by defining a new predicate and folding.
The formal details are in
\cite{LeuschelDeSchreyeDeWaal:JICSLP96,%
GlueckJorgensenMartensSorenson:control,Leuschel:phd,CPD:megapaper}.
On the control side, there are two important issues that
 arise, which we address in the next two subsections.

\subsection{Improved Local Specialisation}

In addition to enabling tupling- and deforestation-like optimisations,
 conjunctive partial deduction also
 solves a problem already identified in \cite{Owen:META88}.
Take for example a metainterpreter containing the
 clause
 {\normalsize $\mathit{solve}(X) \leftarrow \mathit{exp}(X), \mathit{clause}(X,B),$
    $\mathit{solve}(B)$}, where
 {\normalsize $\mathit{exp}(X)$} is an expensive test which for some reason cannot be
 (fully) unfolded.
Here ``classical'' partial deduction faces an unsolvable
 dilemma, e.g., when specialising
 {\normalsize $\mathit{solve}({\bar s})$},
  where ${\bar s}$ is some static input.
Either it unfolds {\normalsize $\mathit{clause}({\bar s},B)$},
 thereby propagating the static input ${\bar s}$
 over to {\normalsize $\mathit{solve}(B)$},
 but at the cost of duplicating {\normalsize $\mathit{exp}({\bar s})$}
 and most probably leading to inefficient programs
 (cf.\ Example~\ref{ex:inboth:duplication}).
Or ``classical'' partial deduction can
  stop the unfolding, but then
 the partial input ${\bar s}$ can no longer be exploited inside
 {\normalsize $\mathit{solve}(B)$} as it will be specialised in isolation.
Using conjunctive
 partial deduction however,
 we can be efficient {\em and\/} propagate information at the same time,
 simply by stopping unfolding and specialising the conjunction $C$ =
 {\normalsize $\mathit{clause}({\bar s},B) \wedge \mathit{solve}(B)$}.
This will result in a specialised clause of the form:
 {\normalsize $\mathit{solve}({\bar s}) \leftarrow \mathit{exp}({\bar s}),
  \mathit{conj\_cs}({\bar
s})$},
 where
 $\mathit{conj\_cs}$ is the predicate defined by the clauses resulting from
 specialising the conjunction $C$.
Experiments in  \cite{JorgensenLeuschelMartens:CPPD,Leuschel:phd})
show that conjunctive partial deduction gives superior specialisation
on programs as the above.

An additional benefit of this is that there
 is now much less need for non-determinate unfolding rules.
For instance, while classical partial deduction with
 (almost) determinate unfolding performs badly on
 highly nondeterministic programs, this is no longer
 true for conjunctive partial deduction.
The following table (extracted from \cite{JorgensenLeuschelMartens:CPPD})
 for the ``contains'' benchmark underlines this:

  \begin{center} 
    \begin{footnotesize}
      \begin{tabular}{c|ccc|c}
         System &  {\sc ecce}  & {\sc ecce}  & {\sc mixtus} & {\sc ecce}  \\
        Type of PD &   Classical  & Classical  & Classical & Conjunctive \\
        Unfolding &    almost determinate  & non-determinate & non-determinate & almost determinate\\
\hline
        Speedup & 1.18  & 11.11 & 6.25 & 9.09\\ 
      \end{tabular}
    \end{footnotesize}
  \end{center}

\subsection{Global Control and Implementation}

Now, while it becomes easier to define an unfolding function that
exploits all available information, there is a
 termination problem specific to conjunctive partial deduction.
It lies in
 the possible appearance of ever growing conjunctions at the global level.
To cope with this, generalisation in the context of conjunctive
 partial deduction must include the ability to {\em split\/}
 a conjunction into several
 parts, thus producing {\em subconjunctions\/} of the original one.
A method to deal with this problem has been developed
 in \cite{GlueckJorgensenMartensSorenson:control,CPD:megapaper},
 which treats 
 the conjunction operator as an associative operator within $\homeo$
 and then splits a conjunction according to the growth detected by
 $\homeo$ and computes the {\em msg} with the best matching
 subconjunction.  
 This splitting reintroduces the problem that no
   information is exchanged between different components of a leaf,
   however, the components are now conjuncts instead of individual
   atoms.
 
For example, if the conjunction $C$ =
 $p(X),$ $ q(f(X),s(0)),$ $ r(f(X)),$ $  s(X)$ has $C'$ =
 $q(Z,0),$ $ r(Z)$ as ancestor, then $C'$ is embedded in $C$ and
 one would split $C$ into
 $C_{1}$=$p(X)$, $C_{2}$=$q(f(X),s(0)),$ $ r(f(X))$,
 $C_{3}$=$s(X)$.
One would then compute the {\em msg}
 of $C'$ and $C_{2}$, giving $C''$ =
 $q(Z,C),r(Z)$ as generalisation.
Finally, as in classical partial deduction, one would then
 specialise $C''$ instead of $C'$.
 
Apart from the above modifications,
 the conventional control notions described
 earlier also apply in a conjunctive setting.
Notably, the concept of characteristic trees can be generalised to
 handle conjunctions.
The {\sc ecce} system \cite{Leuschel96:ecce-dppd}, discussed earlier,
 has been extended to handle conjunctive partial deduction
 and the extensive experiments conducted in
  \cite{JorgensenLeuschelMartens:CPPD,Leuschel:phd}
 suggest that it was possible to
 consolidate partial deduction and unfold/fold
 program transformation, incorporating most of the power of the latter while
 keeping the automatic control and efficiency
 of the former.

\subsection{Relationship to Unfold/Fold}
Unfold/fold
transformations of logic programs
 have been studied by \cite{TamakiSato:iclp84,Pettorossi94:jlp}, and
were originally introduced by
 \cite{BurstallDarlington:ACM77} in functional programming.
The relation between unfold/fold and partial deduction
 has been a matter of
research, discussion, and controversy over the
years
\cite{BossiCoccoDulli:toplas90,Proietti93:jlp,Pettorossi94:jlp,Seki:JLP93,%
  CPD:megapaper}.  
Within the fold/unfold transformation framework, there is work that
aims at developing strategies that can be automated.  For example,
\cite{Pettorossi94:jlp} describe a strategy for partial deduction.
Their technique relies on a simple folding strategy involving no
generalisation, so termination of the strategy is not guaranteed.
Similar approaches are described in
\cite{Proietti:PLILP91,Proietti93:jlp} (in~\cite{Proietti93:jlp}
generalisation is present in the notion of ``minimal foldable upper
portion'' of an unfolding tree).
Also, as unfold/fold transformations are equivalence preserving one
needs a post-processing reachability analysis to delete dead code (for
the queries under consideration).  Such a reachability analysis is
an integral part of  partial deduction algorithms.

Another related approach is described in \cite{Boulanger:JSC93}.  The
authors extend OLDT \cite{TamakiSato:ICLP-86} to cope with
conjunctions, similar to the way conjunctive partial deduction extends
classical partial deduction.  They then use abstract interpretation
(in practice, generalisation is used as in partial deduction) to build
a finite extended OLDT tree from which a specialised program is
extracted.  A major difference with (conjunctive) partial deduction is
that a single global tree is built.  The strategies needed to guide
the construction of the optimal tree are lacking. It is plausible
that the local and global control strategies developed for partial
deduction could be translated into adequate strategies for building
the extended OLDT tree.

In general, unfold/fold (together with a post-processing reachability
analysis) can be seen to subsume both partial deduction and
conjunctive partial deduction.  However, from a practical point of
view, partial deduction has advantages.  Due to its more limited
applicability, and its resulting lower complexity, the transformation
can be more effectively and easily controlled.  In fact, to our
knowledge, no fully automatic unfold/fold systems are available for
experimentation.
However, some explicit strategies for unfold/fold transformation have
 been proposed and recently a semi-automatic system has
 been developed \cite{MAP}. 
Let us consider some of the most well-known strategies:
loop absorption and generalisation (LAG)~\cite{Proietti93:jlp} and
unfold-definition-fold (UDF)~\cite{Proietti:PLILP91} (see also
\cite{Pettorossi94:jlp}). 
Both LAG and UDF use a class of
computation rules, called {\em synchronised descent rules};
 a heuristic tuned towards foldability (and therefore,
 indirectly, termination of the strategy) and the generation of optimal
transformed programs.  However, neither LAG nor UDF
guarantee termination in general.  Instead, classes of
programs are identified for which termination is ensured.  As we have
seen in this article, in partial
 deduction, methods have been proved to secure
termination for {\em all\/} programs. Moreover, notions capturing the
specialisation behaviour, such as characteristic trees, have been
shown instrumental in providing precise generalisation.  This level of
technical detail has facilitated implementation, experimental evaluation
and further improvements.

\subsection{Relationship to other approaches}

\paragraph{Techniques in Functional Programming.}
Partial deduction and related techniques in functional programming are
often very similar \cite{GlueckSorensen:PLILP94} (and
cross-fertilisation has taken place).  Actually, conjunctive partial
deduction has in part been inspired by supercompilation of functional
programming
 \cite{Turchin:toplas86,GlueckSorensen:PE96,SorensenGlueckJones:jfp,%
Sorensen:PE98}
(and by unfold/fold transformation techniques) and the
techniques have a lot in common. However, there are still some subtle
differences.  Notably, while conjunctive partial deduction can perform
deforestation {\em and\/} tupling, supercompilation
  is incapable of
achieving tupling.  On the other hand, the techniques developed for
tupling of functional programs \cite{Chin:PEPM93,Chin:WSA93} are
incapable of performing deforestation.
 
The reason for this extra power conferred by conjunctive
 partial deduction, is that
 conjunctions with shared variables can be used 
 both to elegantly represent {\em nested function calls\/}
\begin{quote}
~~~$f(g(X))$ ~~~~~~~\, $\mapsto$~~~~
 $g(X,\underline{\mathit{ResG}}),$
          $ f(\underline{\mathit{ResG}},\mathit{Res})$
\end{quote}
as well as {\em tuples\/}
\begin{quote}
$\langle f(X), g(X)\rangle$ ~~~~ $\mapsto$~~~~
$g(\underline{X},\mathit{ResG}),$ $ f(\underline{X},\mathit{ResF})$
\end{quote}
or any mixture thereof.  The former enables deforestation while the
latter is vital for tupling, explaining why conjunctive partial
deduction can achieve both.
 
Let us, however, also note that actually achieving the tupling or
deforestation in a logic programming context can be harder.  For
instance, in functional programming we know that for the same function
call we always get the same, unique output.  This is often important
to achieve tupling, as it allows one to replace multiple function
calls by a single call.  For example we can safely transform
$\mathit{fib}(N) + \mathit{fib}(N)$ into {\bf \tt let} $X =
\mathit{fib}(N)$ {\bf \tt in} $X + X$. However, in the context of
logic programming, it is unsafe to transform the corresponding
conjunction $\mathit{fib}(N,R1) \wedge \mathit{fib}(N,R2) \wedge
\mathit{Res}$~$is$~$R1 + R2$ into $\mathit{fib}(N,R) \wedge
\mathit{Res}$~$is$~$R + R$ unless it is proven or declared by the user
that the relation $\mathit{fib/2}$ is functional in its first
argument.  Tupling in logic programming thus often requires one to
establish {\em functionality\/} of the involved predicates.  This can
for instance be done via abstract interpretation (c.f.,
Section~\ref{section:discussion}) or via user declarations that are
assumed to be correct or verified through analysis.
 
Furthermore, in functional programming, function calls cannot {\em fail}
  while predicate calls in logic programming can.
This means that {\em reordering\/} calls in logic programming can
  induce a change in the termination behaviour; something which
  is not a problem in (pure) strict functional programming.
Unfortunately, reordering is often required to achieve deforestation
 or tupling.
This means that to actually achieve deforestation or tupling
 in logic programming one often needs an additional analysis
 to ensure that termination is preserved
 \cite{BossiCoccoEtalle:LOPSTR95,BossiCocco:LOPSTR96}.

\paragraph{Partial evaluation of functional logic programs.}
Functional logic programming \cite{Hanus:JLP94}
 extends both logic and functional programming.
A lot of work has recently been carried out on
 partial deduction of such languages
\cite{AlpuenteFalaschiVidal:ESOP96,%
Alpuente:PEPM97,AlbertEtAl:SAS98,AlpuenteEtAl:TOPLAS98,%
AlbertAlpuenteHanusVidal:LPAR99} (treating languages based
 on narrowing) and
\cite{LafaveGallagher:LOPSTR97} (treating languages based
 on rewriting).
The developed control techniques have been strongly influenced by
 those developed for
 supercompilation of functional programs
 and (conjunctive) partial deduction of logic programs.

\paragraph{Compiling Control.}
Another transformation technique close to both partial deduction and
supercompilation is compiling control
\cite{BruynoogheDeSchreyeKrekels:jlp91}. A major difference with
partial deduction is that the purpose is not to specialise a program
based on the available static input but based on a better computation
rule that reorders the execution of (generate and test) programs by
performing tests as soon as their necessary inputs are available. To
do so, the program is executed using a symbolic input (in fact, using
an abstraction that abstracts ground terms by a ``ground'' symbol and
leaves non-ground terms intact) and builds an initial segment of an
infinite SLD-tree using an {\em oracle\/} to define the optimal
execution order. The oracle either selects an atom for one unfolding
step or for complete execution. In the latter case, the answers of the
execution are abstracted using the ground symbol for ground terms (a
more sophisticated abstraction, performing some generalisation on
non-ground terms is needed in cases where this abstraction does not
lead to a finite number of answers).  The obtained incomplete tree is
similar to the SLDNF-tree of partial deduction in that its nodes are
goal statements. A difference 
with major partial
deduction approaches is that a single global tree is built.  Next,
classes of similar nodes are identified in the tree. The similarity
criterion is based on the selected atom and on the predicate symbols
of the atoms presented in the nodes. Finally, the specialised program
is extracted.  In the context of partial deduction, that extraction
can best be understood, as performing a local unfolding for each class
(again using the oracle to guide the selection of atoms) until a leaf
is reached that is a member of some class. At which point the
resultants can be extracted and give rise to the specialised program.
It is noteworthy that examples are treated which go beyond conjunctive
partial deduction in the sense that goals, conjoined in a new
predicate, can have ---for some predicate symbols--- a varying number
of atoms. The atoms in question are joined in a list structure.


\section{Discussion and Conclusion}
\label{section:discussion}

\subsubsection*{Research Challenges}

Despite over 10 years of research on logic program
 specialisation, 
 there are still plenty of research challenges related to
 improving the actual specialisation capabilities.
Below, we present what we believe to be the major research challenges
 for the coming years.

\paragraph{Control: Low-level cost model.}
Existing systems do not use a sufficiently precise model of the
compiler of the target system to guide their decisions during
specialisation.  We have seen that determinate unfolding will usually
prevent drastic slowdowns, but it is unable to exclude all slowdowns.
Moreover, it is sometimes too conservative and prevents important
improvements.  While there is some recent work \cite{Debray:PEPM97} to
address this, it is a largely ignored area and some of the problematic
issues raised in \cite{VenkenDemoen:NGC88} are still valid today.

 A suitable
 {\em low-level cost model\/}
  would allow a partial deduction system to make more informed
choices about the local control (e.g., is this unfolding step going to be detrimental to
 performance) and global control (e.g., does this extra polyvariance
 really pay off).
However, such a low-level cost model will depend on both the particular
 Prolog compiler and on the target architecture and it is hence
unlikely that one can find
an appropriate 
mathematical theory.
This means that further progress on the control of partial deduction will 
 probably not come
 from ever more refined mathematical techniques such as new wqos, but 
 probably more
 from heuristics and {\em artificial intelligence\/} techniques such
 as case-based reasoning or
 machine learning.
For example, one might imagine a
 {\em self-tuning\/} system, which derives its own cost model of the
 particular compiler and
 architecture by trial and error.
Such an approach has already proven to be highly successful in
 the context of optimising scientific linear algebra software
 \cite{whaley00:ATLAS}.
Some promising initial work on cost models for logic and functional programming
 has already been made in \cite{AlbertEtal:LOPSTR2000,AlbertEtal:LOPSTR2001}.

\paragraph{Predictable specialisation.}
Another drawback of existing specialisation systems
 (especially for online systems) is the lack
 of predictability for both the specialisation time and for the size
 of the generated residual program.

Indeed, while existing online
 systems and methods guarantee termination, their use sometimes 
results in code explosion without achieving substantial specialisation.
One situation where this tends to happen
 is when the program to be specialised
  has a combination of
  arguments that can grow and shrink and when
  the initial atom to be specialised has
  partially instantiated parameters.
The problem is that techniques such as $\homeo$
  have, even given a fixed initial atom,
  no upper bound on the length of admissible
  sequences.
For example, 
 $\langle p(a,b), p(f(b),g(f(b),f(a)))\rangle$ is admissible wrt
  $\homeo$, as the growth of the second argument
  has been countered by the first argument
   (where we have $a\nothomeo f(b)$).
A good example where such a behaviour can appear
 during specialisation is the
 ``groundunify'' benchmark within the
  {\sc dppd} library
  \cite{Leuschel96:ecce-dppd},
  where two arguments are the terms to be unified
  (which are decomposed and thus usually shrink during
   specialisation)
   and another argument is the unifier
  so far (which will usually grow during specialisation).
Using determinate unfolding for local control
 and $\homeo$ and characteristic trees for global control
  will lead to a global
 tree with 480 nodes and 85 specialised predicate
  definitions for this benchmark.
The specialisation effort here is out of proportion
 with the actual speedup obtained.

Developing control techniques with predictable and reasonable
 specialisation complexity is thus a worthwhile, but also challenging
 research objective.
Alternatively, developing an
{\em incremental partial deduction\/} approach
 could overcome these problems in some cases.
Indeed, one could start by a very conservative partial deduction and then
 incrementally adapt the partial deduction, concentrating the
 efforts on the parts where improvements in efficiency or precision will arise.
This could go hand-in-hand with a self-tuning system and a low-level
 cost model.
Finally, as a side-benefit a user could stop the partial deduction at any point and 
 still obtain a correct specialised program.

\paragraph{Improved precision: Combining program specialisation and abstract
interpretation.}
   As we have seen, $\homeo$ and characteristic trees
 provide a quite refined way to decide when
 the generalisation has to be applied.
However, once a growth has been detected by $\homeo$, all of
 these existing specialisation techniques still rely on
 rather crude generalisation functions, such as
  $\mathit{msg}$, because the
 resulting generalisation has to be expressed as
  an atom, which implicitly represents all its instances.
For instance, if we add the atom
 $A_2 = p(f(a))$ as a child of
 $A_1 = p(a)$ in a global tree
  then the
 homeomorphic embedding $\homeo$ will signal danger ($A_1\homeo A_2$)
 and one can even pinpoint the extra $f(.)$ in $A_2$ as the potential source
 of non-termination.
But the $\mathit{msg}$ of $A_1$ and $A_2$---the most specific expression
 which is more general than both $A_1$ and $A_2$---is just
 $p(X)$ and no use of the
information provided
 by $\homeo$ was made (nor is it possible to do so in classical
  partial deduction).
In particular, atoms like $p(b)$ and $p(g(a))$ are also instances
 of $p(X)$, possibly leading to unacceptable losses of precision.
In some cases the characteristic tree based global control will
 avoid these imprecisions.
However, the present generalisation operation on the characteristic
 trees themselves is still a bit crude (common initial
 subsection).
 We think this problem in particular
 and other precision problems in general
 can be overcome by providing a better
 integration
  of partial deduction with abstract interpretation.
 This will also add other benefits, such as
  bottom-up success information propagation and success information
  propagation between atoms at the global level as well.
 
A full integration of partial deduction with {\em abstract interpretation\/} 
is thus another
 of the big challenges.
Indeed, it is often felt that there is a close relationship between
 abstract interpretation and program specialisation. 
Some techniques
 preceding the recent advancements of partial deduction, notably
 compiling control \cite{BruynoogheDeSchreyeKrekels:jlp91} and  the work in
 \cite{Boulanger:JSC93} combine features of abstract interpretation with features
 of partial deduction.
 Recently,
 there has been a lot of interest
 in the integration of these two techniques 
\cite{Jones:LLC94,%
LeuschelDeSchreye:PLILP96,PueblaHermenegildo:LOPSTR96,Jones:SAS97,%
PueblaGallagherHermenegildo:SP97,Leuschel:JICSLP98,%
GallagherPeralta:PEPM00}.
The use of more refined abstract domains,
  improved bottom-up and side-ways information propagation, will
 improve specialisation and precision and opens up new areas for
 practical applications, such
 as  {\em infinite model checking\/}
 \cite{LeuschelMassart:LOPSTR99,LeuschelLehmann:Coverability,%
FioravantiEtAl:VCL01}.
In fact, such a combined approach enables optimisations (and analysis) which
 cannot be achieved by either method alone
  \cite{LeuschelDeSchreye:PLILP96}.
Finally, having more precise generalisation capabilities
 might actually make the global and local control of
 partial deduction simpler, as much less precision would
 be lost if the control makes a ``wrong'' decision.

\paragraph{Tabling and constraints.}
Finally, features such as co-routining, constraints, and tabling
provided by the latest generation Prolog systems, apart from being
very useful in practice, also mean that declarative programming is now
much more of a reality than in a classical Prolog environment.  It is
thus important that partial deduction be adapted to treat these
features.

First, logic programming with
 inequality constraints
 provides a more sophisticated way to handle
 negated literals: by using so called
 {\em constructive negation\/} one can
 even specialise non-ground negative literals
\cite{Chan:META88}.
This idea was successfully used within
 the {\sc sage} system \cite{Gurr:PHD}.

On the side of specialising arbitrary constraint logic programs
 themselves, we can mention the works of
\cite{Smith:naclp90,Smith:pepm91,MarriottStuckey:POPL93,%
EtalleGabbrielli:TCS96,BensaouEtAl:TCS98}.
Future work should 
advance the state of the art of specialising constraint logic
programming to that for standard logic programming.
First steps in that direction have been presented in
\cite{FioravantiEtAl:Woid99,FioravantiEtAl:Lopstr00}.

In the context of
  tabled-evaluation of logic programs \cite{ChenWarren:JACM96},
 some specialisation
 techniques have been successfully built
 into the execution mechanism itself
\cite{UnifFact:POPL95}, but there has been relatively
 little work on transforming or specialising tabled logic
 programs.
Somewhat surprisingly, as shown in
\cite{LeuschelMartensSagonas:xsb,SagonasLeuschel:ACMCS98},
tabled logic programming
 generates some new challenges to program transformation
 in general and partial deduction in particular.
For example, contrary to the untabled setting,
 unfolding can transform a program terminating under tabled-evaluation
  into program that is non-terminating under tabled-evaluation.

\subsubsection*{Practical Challenges: On the uptake of partial deduction}
Despite some success stories and the increasing integration of
 partial deduction methods into compilers (e.g., the Mercury
 compiler specialises higher-order predicates such as map), the
 general uptake of partial deduction methods might be deemed
 disappointing.
In the following we present some factors which we believe
 explain this situation:
 
\begin{zitemize}
    \item non-declarative features:
  most Prolog programs contain some form of non-declarative parts.
 Now, whereas systems such as {\sc mixtus} or {\sc paddy}
  can handle such programs, non-declarative features
  impose severe restrictions on the specialiser, and the
   speedups obtained are often disappointing.
 In addition, most programs do not have a clear distinction
 between pure and impure parts, and it is thus difficult to apply
 systems such as {\sc sp} or {\sc ecce} 
 to large parts of the code.

To solve this problem, one might turn to
 more powerful, complementary analysis techniques,
 so as to lift some of the restrictions in the
 presence of impure features. E.g., one might
 integrate a partial evaluation
system into Ciao
Prolog where it could benefit from other analyses and/or optional
user declarations.
However, this is likely to involve considerable research
 and development effort.

Another solution is to promote a more declarative style of programming,
 more suitable for specialisation: e.g., programs written in
  Mercury,  G\"{o}del, or even pure Prolog with
  declarative built-ins and if-then-else and clearly
   separated i/o (or ``declarative'' i/o).

\item  For the offline approach,
 the lack of an implementation with a fully automatic bta,
 means that basically
 only expert users can use the current systems.
 However, as discussed earlier, 
some important steps towards automatisation of bta have recently been
made and hopefully, they will soon become part of available systems.

\item In 
 principle,  existing online systems such as {\sc mixtus} and
 {\sc ecce} are fully automatic and can be used by a na\"{\i}ve user.
 However, as we have discussed above,
  for more involved programs, these systems
   can sometimes still lead to substantial code explosion and
 substantial specialisation times.
 Currently, to overcome this, user expertise is still required to fine tune the
 specialisation of the program at hand.

\item
  Also, as we have seen above,
 existing systems do not use a sufficiently
  precise low-level cost model to guide the specialisation
 process.  Consequently, they are unable to
  exclude anomalies such as slow-down of the specialised
  program. 

  \item Finally, existing specialisers are not yet
 fully integrated within a
programming environment.
On the one hand, this means that it is more cumbersome to apply these
tools (the user has to link up the specialised code with the rest
 of his application, the user has to know when parts of his application
 have to be respecialised,...).
On the other hand, this means that currently specialisers are often
 only applied late in the development on
   already hand-optimised code.
This makes the specialisers task more difficult and
  reduces the speedup and benefit.
  
 Thus, one of the practical challenges is to
  produce a partial deduction system that is fully integrated with a compiler,
 so that it can  be easily used during and as part of
  the development process.
Also, provide support for non-declarative parts and modules.
Another difficulty is the interference with debugging, as
 users want to debug the code they wrote, not the specialised code.
\end{zitemize}

However, we feel that it is possible to overcome the above
 obstacles and that in the not too distant future one could
 lift program specialisation towards more widespread practical use
 and realise its potential as a
 tool for systematic program development.
As to the future of the off-line versus on-line debate,
 we believe that hybrid approaches might prove to be
 the way to go for many applications, delivering a good
 compromise between fast
 transformation speeds and precise specialisation.
In fact, one approach which we have already found to be useful
 \cite{LeuschelLehmann:Coverability} is to first perform
 an off-line specialisation followed by an on-line
 specialisation.

\subsection*{Acknowledgements}

First, we would like to thank
 Danny De Schreye, Andr\'{e} de
Waal, Robert Gl\"{u}ck, Bern Martens, and Morten Heine 
S{\o}rensen for all the joint work and stimulating discussions.
(Actually, some sentences from our joint works
   have probably made it into the paper.)
We are grateful to Bart Demoen for his valuable feedback and insights,
especially on Prolog performance issues.
The authors would also like to thank the LOPSTR community for interest,
 enlightening comments, and discussions on many of the subjects
 covered in this paper.
Finally, we would like to thank the anonymous referees for their
 very extensive and thorough feedback, which substantially
 helped us to improve paper.
 


\end{document}